\documentclass[sigconf,nonacm]{acmart}
\usepackage{algorithm} 
\usepackage[utf8]{inputenc}
\usepackage[T1]{fontenc}
\usepackage{textcomp}

\AtBeginDocument{%
  }

\newcommand{\xj}[1]{#1}

\newcommand{\tl}[1]{#1}

\renewcommand\footnotetextcopyrightpermission[1]{} 
\begin{document}

\title{KeySense: LLM-Powered Hands-Down, Ten-Finger Typing on Commodity Touchscreens}


\author{Tony Li}
\email{haolili@cs.stonybrook.edu}
\affiliation{%
  \department{Department of Computer Science}
  \institution{Stony Brook University}
  \city{Stony Brook}
  \state{New York}
  \country{USA}
}

\author{Yan Ma}
\email{yan.ma@kean.edu}
\affiliation{%
  \department{Computer Science Department}
  \institution{Kean University}
  \city{Union}
  \state{New Jersey}
  \country{USA}
}

\author{Zhuojun Li}
\email{lizj23@mails.tsinghua.edu.cn}
\affiliation{%
  \department{Department of Computer Science and Technology}
  \institution{Tsinghua University}
  \city{Beijing}
  \country{China}
}

\author{Chun Yu}
\email{chunyu@tsinghua.edu.cn}
\affiliation{%
  \department{Department of Computer Science and Technology}
  \institution{Tsinghua University}
  \city{Beijing}
  \country{China}
}

\author{IV Ramakrishnan}
\email{ram@cs.stonybrook.edu}
\affiliation{%
  \department{Computer Science}
  \institution{Stony Brook University}
  \city{Stony Brook}
  \state{New York}
  \country{USA}
}

\author{Xiaojun Bi}
\email{xiaojun@cs.stonybrook.edu}
\affiliation{%
  \department{Department of Computer Science}
  \institution{Stony Brook University}
  \city{Stony Brook}
  \state{New York}
  \country{USA}
}

\renewcommand{\shortauthors}{Li et al.}

\begin{abstract}

\tl{Existing touchscreen software keyboards prevent users from resting their hands, forcing slow and fatiguing index-finger tapping (“chicken typing”) instead of familiar hands-down ten-finger typing. We present KeySense, a purely software solution that preserves physical keyboard motor skills.} KeySense isolates intentional taps from resting-finger noise with cognitive–motor timing patterns, and then uses a fine-tuned LLM decoder to turn the resulting noisy letter sequence into the intended word. \tl{In controlled component tests, this decoder substantially outperforms 2 statistical baselines (top-1 accuracy 84.8\% vs 75.7\% and 79.3\%).} A 12-participant study shows clear ergonomic and performance benefits: compared with the conventional hover-style keyboard, users rated KeySense as markedly less physically demanding (NASA-TLX median 1.5 vs 4.0), and after brief practice, typed significantly faster (WPM 28.3 vs 26.2, p <0.01). \tl{These results indicate that KeySense enables accurate, efficient and comfortable ten-finger text entry on commodity touchscreens, without any extra hardware.}

\end{abstract}



\begin{teaserfigure}
  \includegraphics[width=\textwidth]{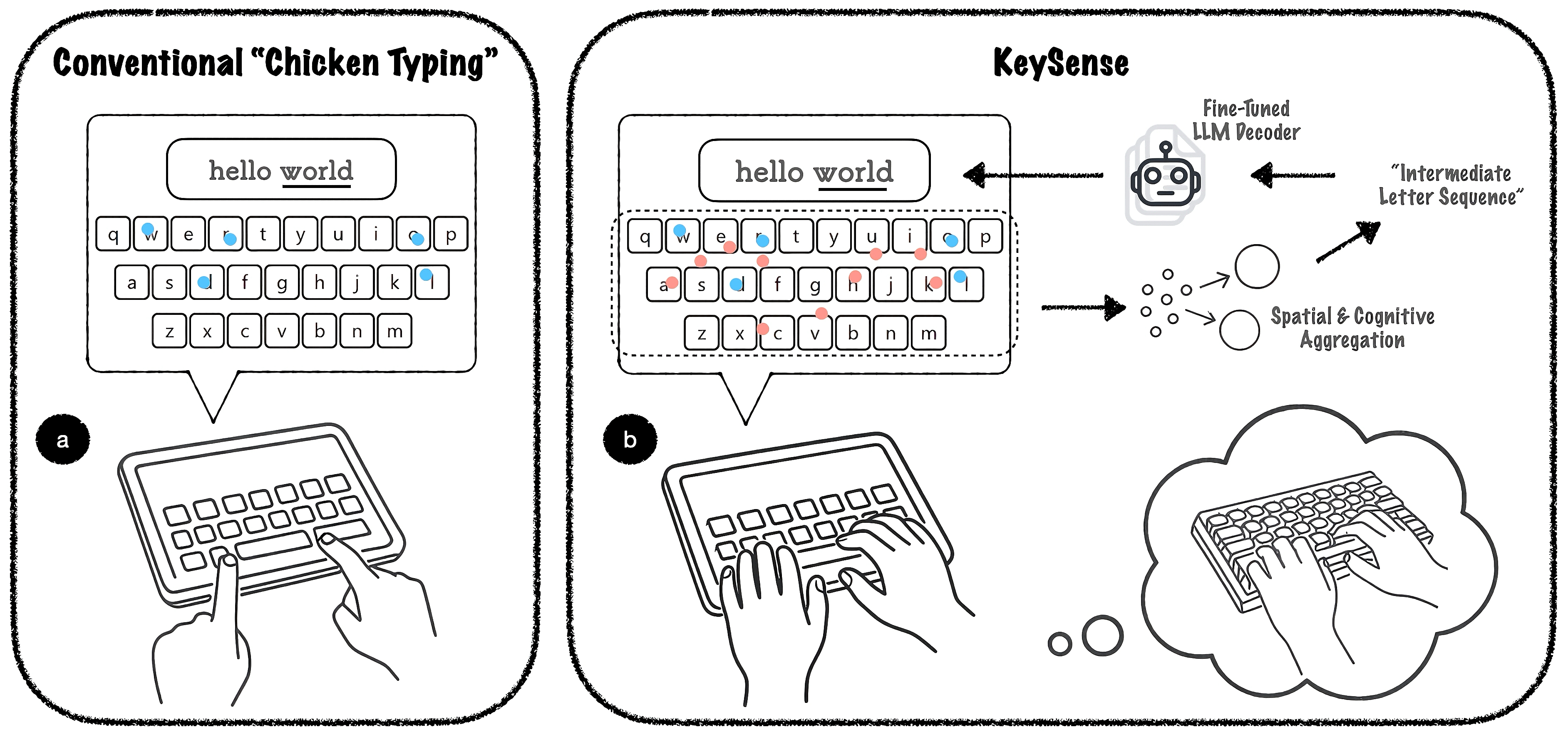}
  \caption{KeySense enables hands-down, ten-finger typing on commodity touchscreens by letting users rest their fingers while the system leverages an LLM to decode the intended word. 
(a) Conventional touch keyboards encourage two-finger "chicken typing" where users hover. 
(b) With KeySense, users keep both hands down; blue dots mark intentional reaches, red dots mark resting touches, and the underlined text shows the current word in progress.}
  \label{fig:teaser}
  \Description{Teaser comparing “chicken typing” with KeySense on a commodity tablet. Panel (a): users peck with two fingers and depend on spell-check. Panel (b): both hands rest on the screen; blue dots = intentional reaches, red dots = resting contacts; the current word is underlined. The system leverages a fine-tuned LLM to turn these touches into the intended word.}
\end{teaserfigure}


\maketitle

\section{Introduction}

\tl{Ten-finger typing on physical keyboards is fast, comfortable, and built on decades of motor learning: all fingers rest on a home row, and keystrokes are produced through small reaches. On touchscreens, however, this motor program breaks down. Because any surface contact risks triggering an unintended keystroke, users must hover their hands and peck with one or two fingers (“chicken typing”; Figure~\ref{fig:teaser}a), leading to fatigue and preventing transfer of motor skills.}

\xj{This difficulty reflects a deeper issue articulated by Buxton’s three-state model of input~\cite{buxton1990three}. Whereas some pointer devices such as mice can cleanly separate out-of-range, tracking, and dragging states, touchscreens collapse tracking and dragging states: any contact with the screen is treated as an intentional input, producing nearly identical $(x,y,t)$ signals. This phenomenon also known as the MIDAS-touch problem~\cite{jacob1995eye}. This loss of state separation is the fundamental reason why hands-down typing on touchscreens is inherently ambiguous.}


Prior approaches have explored specialized hardware such as pressure sensors, depth cameras, or instrumented surfaces \cite{gu2021typeboard,li2023restype,zhang2022typeanywhere} \xj{to separate accidental contacts from intentional touch input,} but such components are absent from commodity tablets. Software-only methods have instead focused on sentence-level decoding or simplified heuristics \cite{kim2013tapboard,kim2016tapboard,vertanen2015velocitap}, which do not resolve the ambiguity introduced by resting fingers in hands-down input. The central challenge remains: with only raw $(x,y,t)$ touch events, can intentional keystrokes and incidental contact be distinguished?


\tl{We present \emph{KeySense}, a software-only system that enables comfortable, hands-down ten-finger typing on commodity, pressure-insensitive touchscreens. The system consists of two stages.} A lightweight front end reconstructs touch \emph{threads}, groups near-synchronous contacts within a short cognitive window, and selects one representative per cluster using a reach-aware heuristic that suppresses resting touches. The resulting noisy letter sequence is then corrected by a compact LLM (FLAN–T5-small) fine-tuned on a synthetic corpus that models human surface-typing errors.

Our evaluations show that this combination of structured pre-processing and LLM-based decoding yields strong accuracy. On a synthetic test set, the fine-tuned FLAN–T5-small achieves \textbf{84.8\%} Top-1 accuracy, substantially outperforming two statistical baselines (\textbf{75.7\%} and \textbf{79.3\%}) and zero-shot LLM variants.

A 12-participant study further demonstrates the practical impact of this approach. Participants rated KeySense as far less physically demanding than the hover-typing baseline (NASA-TLX median 1.5 vs.\ 4.0), and after a brief learning period they typed significantly faster with KeySense in the final session (28.3 vs.\ 26.2 WPM; $F(1,118.71)=8.58, p=0.0041$). These results show that by decoding noisy multi-touch input rather than suppressing it, commodity touchscreens can support a comfortable ten-finger typing style that ultimately surpasses conventional interaction.

\tl{Taken together, this work shows that a purely software-based approach can restore the motor and ergonomic benefits of ten-finger typing on commodity touchscreens. Beyond demonstrating a practical system, the paper also provides the first detailed characterization of hands-down typing behavior on passive surfaces, highlighting how resting and reaching patterns emerge as users adapt to this interaction style.}
\section{Related Work}

As an essential component of Human–Computer Interaction, text entry has been extensively studied in the past few decades. This section discusses two key areas on which this work is built: ten-finger typing and tap-typing decoders.

\subsection{Ten-Finger Typing}
\label{subsec:tenfinger}

Ten-finger typing preserves bimanual parallelism and transfers decades of QWERTY muscle memory to touch surfaces, enabling higher sustained throughput with lower visual demand than single- or two-finger methods, especially on tablet-sized displays where hands can rest naturally. Against this backdrop, work on ten-finger text entry without physical keycaps has developed along three complementary fronts: characterizing how experts type on glass, modeling and adapting the keyboard to users, and improving separability between intentional taps and incidental contacts. 

Foundational empirical studies showed that expert touch patterns on flat glass are systematic and thus learnable despite the absence of tactile landmarks \cite{findlater2011typing}. Building on this, personalized spatial models that adapt key-target distributions to each individual demonstrably reduce errors and established personalization \cite{findlater2012personalized}.

\tl{A major challenge for ten-finger posture is the resting contacts that confound naive tap detection. \emph{Kim et al.} proposed \textit{TapBoard}, utilizing a distinction between brief, low-motion keystrokes and longer, stable rests \cite{kim2013tapboard}. \emph{Kim and Lee} later introduced \textit{TapBoard 2}, streamlining the interaction so users could fluidly switch between typing and pointer-like manipulation on the same surface \cite{kim2016tapboard}. Complementary explorations augmented QWERTY touch keyboards with multi-touch gestures for non-alphanumeric input \cite{findlater2012beyond} and compressed the full layout into a single horizontal line while preserving familiar spatial ordering \cite{li20111line}. Together, these systems showed that simple temporal and kinematic cues and layout/gesture design, can suppress noise but also impose behavioral constraints on how users must “tap.”}

A second line of work adds sensing. \tl{Pressure-sensitive keyboards such as \textit{TypeBoard} identify unintentional touches \cite{gu2021typeboard}, while \textit{T-Force} extends this idea to a three-state virtual keyboard that leverages typing force to distinguish touch states \cite{faleel2023t}. Building on \textit{TypeBoard}, \textit{ResType} further explored adaptive and invisible layouts on pressure-sensitive hardware \cite{li2023restype}.} Wearable-augmented systems attach lightweight sensors to fingers and decode QWERTY input on arbitrary surfaces \cite{streli2022taptype,zhang2022typeanywhere}, and vision-based approaches infer keystrokes from hand tracking, showing that surface typing can be decoded without capacitive touch \cite{richardson2020decoding,roeber2003typing}. These sensing-rich approaches push performance and flexibility but raise hardware and deployment issues that limit adoption on commodity tablets.

On the decoding side, probabilistic inference has made “invisible keyboards” feasible even on ordinary flat surfaces. \emph{Shi et al.} introduced \textit{TOAST}, which learns user-anchored latent layouts and combines spatial–temporal evidence with lexical priors to recover word sequences with competitive accuracy and speed in eyes-free scenarios \cite{shi2018toast}. At the extreme, mid-air methods reconstruct ten-finger sequences from 3D hand tracking, revealing the capacity of expert motor plans once contact ambiguity is removed \cite{yi2015atk}. These results collectively argue that expert ten-finger behavior is modelable, that adaptation pays off, and that principled inference can overcome much of the ambiguity inherent in on-glass typing.

\begin{table*}[h]
\centering
\begin{tabular}{l|lll}
\toprule
\textbf{System} & \textbf{Hardware Requirement} & \textbf{Primary Innovation} & \textbf{Decoder Type} \\
\midrule
\textbf{KeySense (Our work)} & \textbf{Commodity Touchscreen} & Heuristic Filtering + LLM Decoding & Fine-tuned LLM \\
TypeBoard \cite{gu2021typeboard} & Pressure-sensitive Surface & Pressure-based Touch Prevention & Statistical \\
ResType \cite{li2023restype} & Pressure-sensitive Surface & Adaptive Layout & Statistical \\
\bottomrule
\end{tabular}
\caption{Design space of systems enabling a \emph{hands-down}, ten-finger typing posture. While TypeBoard and ResType validated this interaction paradigm on pressure-sensitive hardware, KeySense is the first to achieve it on commodity touchscreens through a purely software-based, LLM-powered decoding approach.}
\label{tab:related_work}
\Description{Design-space comparison of three hands-down, ten-finger typing systems: KeySense, TypeBoard, and ResType. Columns describe each system's hardware requirement, primary innovation, and decoder type. KeySense uses a commodity touchscreen and a fine-tuned LLM decoder. TypeBoard and ResType both require a pressure-sensitive surface and use statistical decoders, but differ in their primary innovation (pressure-based prevention vs. adaptive layouts and a public corpus).}
\end{table*}

\tl{Within this broader landscape, KeySense targets the most practical and widespread scenario: enabling an ergonomic, \emph{hands-down, ten-finger posture} directly on an \emph{unmodified, pressure-insensitive touchscreen}.} This posture, where users can comfortably rest their fingers as they would on a physical keyboard, is key to reducing fatigue and leveraging existing muscle memory. Within this context, Table~\ref{tab:related_work} systematically compares our approach to the key prior systems that have also successfully enabled a hands-down posture: \textit{TypeBoard}~\cite{gu2021typeboard} and \textit{ResType}~\cite{li2023restype}. \tl{KeySense shares the same interaction goal but makes the crucial leap to commodity, pressure-insensitive hardware.}

\subsection{Tap-Typing Decoders}
\label{subsec:tap-decoders}

Tap typing has long relied on probabilistic decoders that fuse a spatial touch model with a language model to resolve noisy taps. A representative family of systems models per-key touch likelihoods (often Gaussian or learned from user data) and searches for the most probable string under a character or word language model. \tl{\emph{Vertanen et al.} proposed this paradigm at the \emph{sentence} level with \textit{VelociTap}, showing that deferring commitment and decoding whole phrases boost speed and robustness \cite{vertanen2015velocitap}. On small displays, \textit{WatchWriter} adapts statistical decoding to smartwatch and corrects “fat-finger” taps \cite{gordon2016watchwriter}. Prior work demonstrated that strong language priors can render motor-memory–driven typing practical even when key borders are not visually shown \cite{zhu2018typing}.} More broadly,  the decoder itself can be framed as Bayesian inference over touch uncertainty, exploring learned touch models and user-controlled uncertainty (e.g., via pressure) \cite{weir2014uncertain}, while follow-up studies quantified how expanding context from single words to multiple words and sentences further improves decoder performance \cite{vertanen2018impact,vertanen2019velociwatch}.

Beyond direct finger taps, several tap-like modalities alter the sensing or disambiguation channel. \textit{GlanceWriter} uses gaze dwell as a “tap” over letter regions and decodes text, enabling writing by rapidly glancing across letters \cite{cui2023glancewriter}. \textit{TapGazer} couples ambiguous multi-finger tapping with gaze-directed word selection and offloads final disambiguation to the user’s eyes \cite{he2022tapgazer}. \tl{In mixed reality, \textit{TouchInsight} leverages egocentric vision to estimate tap intent with uncertainty-aware inference, allowing rapid touches that are resolved with linguistic context \cite{streli2024touchinsight}.} Outside capacitive sensing, \textit{I-Keyboard} demonstrates a fully imaginary on-surface keyboard with a deep neural decoder that maps touch patterns to characters without rendering keys \cite{kim2019keyboard}, and \textit{Nomon} combines a probabilistic selection model with high-order language models to achieve competitive efficiency in extreme access scenarios \cite{bonaker2022performance}.

Neural and LLM-based decoders have begun to directly rank or generate intended text from noisy taps under strong priors over words and phrases.
\tl{It is demonstrated that compact seq2seq models can jointly correct and complete noisy mobile keyboard input \cite{ghosh2017neural}.}
\tl{\textit{Type, Then Correct} learns to correct mobile keyboard input errors using post-hoc neural inference rather than real-time decoding \cite{zhang2019type}.} \emph{Ma et al.} proposed an LLM-powered smartphone decoder that unifies tap and flexible typing modes and integrates spatial evidence with powerful contextual predictions, and further showed that the same decoder can be applied to virtual reality text entry \cite{ma2025llm,ma2025llmvr}. \tl{Transformer-based language models can dramatically reduce keystrokes by expanding minimal tap input (e.g., initials) into full phrases \cite{cai2023using}. \textit{SkipWriter} also expands abbreviated handwriting into full phrases on tablets \cite{xu2024skipwriter}.} Intelligent post-hoc correction systems such as \textit{JustCorrect} apply semantic cues to repair earlier tap errors with minimal user effort \cite{cui2020justcorrect}.

Taken together, this literature establishes that combining spatial uncertainty with linguistic priors, from word to sentence and now to neural and LLM decoders, substantially improves tap typing \cite{weir2014uncertain,vertanen2015velocitap,vertanen2018impact,vertanen2019velociwatch,ma2025llm}. \tl{Our contribution addresses ten-finger typing on \emph{unmodified} touchscreens by focusing on temporal disambiguation of near-synchronous taps and software-only decoding, complementing prior spatial–linguistic decoders.}
\begin{figure*}[t]
  \centering
  \includegraphics[width=\textwidth]{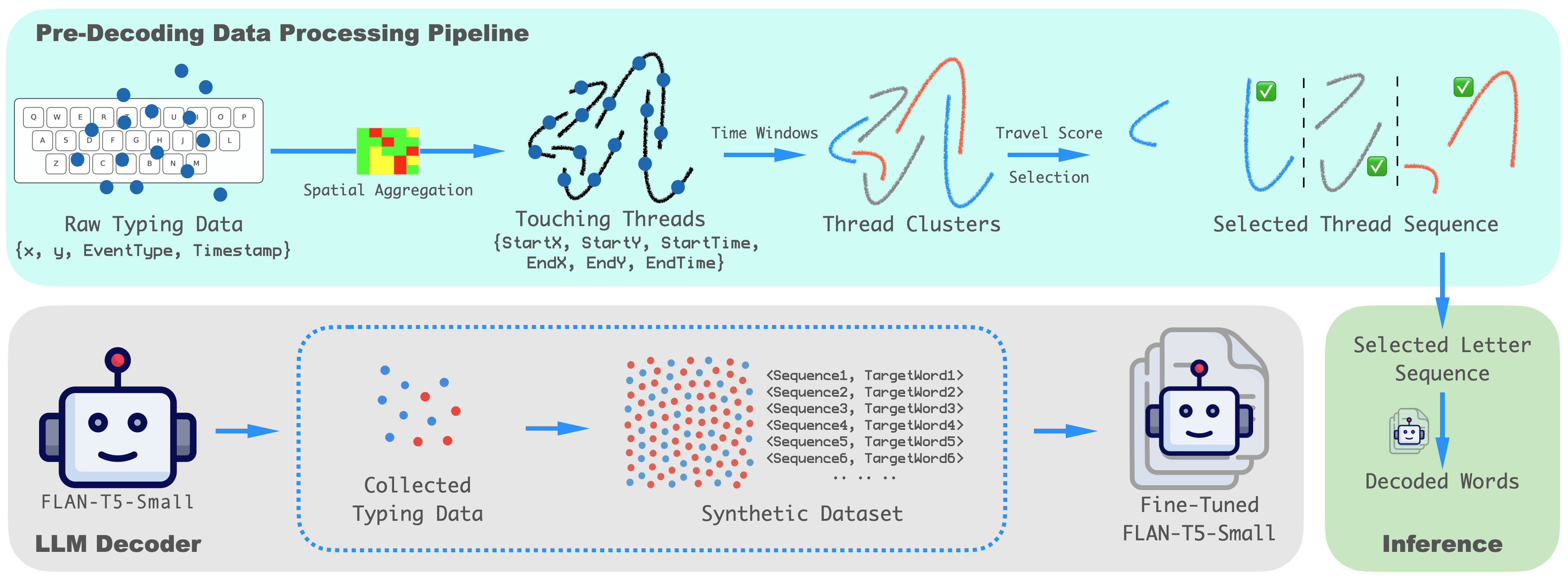}
  \caption{End-to-end overview. \textbf{Top (pre-decoding)}: raw touch events are aggregated into \emph{touch threads}, grouped into \emph{time clusters}, and resolved by a reach-sensitive \emph{travel
  score} to yield one representative per cluster. Each representative is mapped to a key, producing a letter sequence. \textbf{Bottom
  (LLM decoder)}: we parameterize a synthetic corpus from the
  collected typing logs, fine-tune \emph{FLAN-T5-small} on these pairs, and at
  inference decode the letter sequence into a word.}
  \label{fig:pipeline}
  \Description{Pipeline overview from touch input to a corrected word. The figure is organized into two horizontal stages. Top stage (pre-decoding): raw touch events are first grouped into touch threads; threads whose onsets fall within a ~100 ms cognitive window are then clustered; from each cluster, a single representative touch is selected based on travel distance; each selected touch is mapped to its nearest key, producing a provisional letter sequence. Bottom stage (decoding): the provisional letter sequence is fed into a fine-tuned FLAN-T5-small model, which outputs the final corrected word.}
\end{figure*}

\section{Pre-Decoding Data Processing Pipeline}

Figure~\ref{fig:pipeline} summarizes our pipeline and how the
components fit together. The upper track converts noisy, low-level touch events
into a clean, temporally ordered sequence of key hypotheses; the lower track
shows how we train a compact LLM to map that sequence to the intended word.

\noindent\textbf{From events to a sequence (top row).}
Starting with raw points \((x,y,\textit{type},t)\), we (1) aggregate nearby
down/move/up samples into \emph{touch threads},
(2) group threads into \emph{clusters} if their onsets fall within a short
cognitive window in
Section~\ref{subsec:study-intent-gap}), and (3) select one
representative per cluster using a reach-sensitive \emph{travel score}
(Section~\ref{subsec:selection-travel}). Mapping each selected thread to the
nearest key center yields a provisional letter sequence that preserves typing
order while suppressing incidental contacts.

\noindent\textbf{From sequence to word (bottom row).}
We collect typing logs (Section~\ref{subsec:collection}), synthesize a
length-balanced corpus of misspelled–gold pairs
(Section~\ref{subsec:synth}), and fine-tune \emph{FLAN-T5-small} to correct a
single word from its noisy letter sequence (Section~\ref{subsec:fine-tune}).
At inference, the fine-tuned model decodes the letter sequence into
the final word.

\subsection{Cognitive Hypothesis on Typing Interval}
\label{subsec:cognitive-interval}

We consider the setting where one finger executes an \emph{intentional} press while other resting fingers may produce nearly synchronous, \emph{unintentional} contacts. \tl{Our central hypothesis is that human processing limits impose a minimum temporal separation between two \emph{independent} intentional key activations: within a short window \(\tau_c\), co-occurring contacts are overwhelmingly likely to reflect a single planned action accompanied by incidental co-contacts.
Prior work has similarly addressed grouping near-simultaneous touches: \textit{Evans et al.} distinguish multi-user touch identities on tabletops by statistically modeling touch pairs occurring within short temporal windows \cite{evans2017group}.
}


\tl{This view is supported by classic human performance models that place cognitive and motor processor cycles on the order of \(50\)–\(100\)\,ms \cite{card1986model}, and by dual-task research on the \emph{Psychological Refractory Period} (PRP), which shows a central bottleneck that serializes response selection when two responses are scheduled in rapid succession \cite{allen1998psychological}.} Large-scale typing studies report typical inter-key intervals near \(200\)–\(250\)\,ms with only a thin lower tail around \(60\)–\(80\)\,ms, and sustaining rates above \(10\) keys/s (\(<100\)\,ms per keystroke) is difficult without marked errors \cite{feit2016we}. From a perception standpoint, a window of roughly \(100\)\,ms is a canonical bound for events to appear near-simultaneous in interaction \cite{nielsen1993response}.


\tl{These strands suggest a compact interval of roughly \(100\)\,ms where launching two independent intentional keystrokes is highly implausible. We therefore adopt a cognitive window \(\tau_c=100\)\,ms: touches whose onsets fall within this gap are treated as near-synchronous, while touches separated by more than \(\tau_c\) are treated as sequential keystrokes.}

\subsection{Study on Intentional Touch Interval}
\label{subsec:study-intent-gap}

\begin{figure*}[t]
  \centering
  \includegraphics[width=0.8\linewidth]{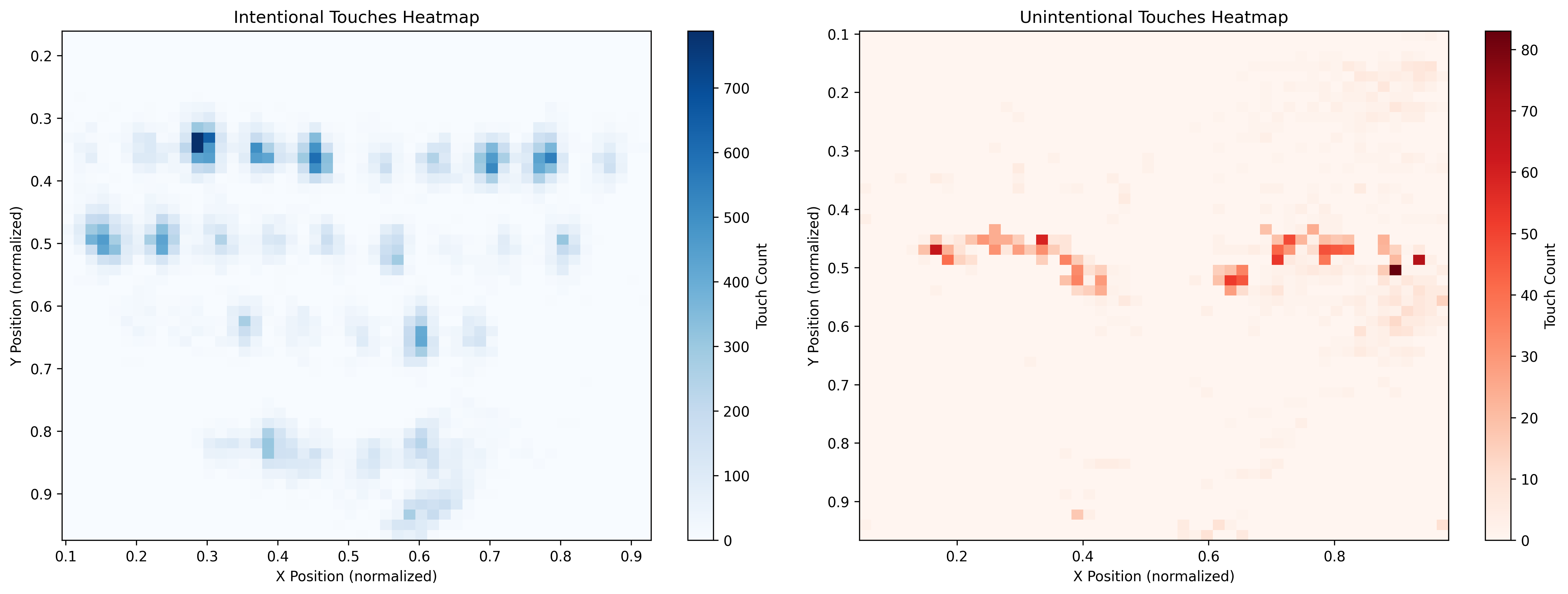}
  \caption{Touch distributions from the ResType dataset \cite{li2023restype}. Intentional taps concentrate over key centers (left), whereas unintentional contacts follow a resting-hand arc (right).}
  \label{fig:touch-heatmaps}
  \Description{Spatial distributions separate intentional taps from incidental rests. Intentional taps concentrate over key centers (x–y plane), while unintentional contacts form an arc consistent with resting fingers/palm.}
\end{figure*}

\tl{Motivated by this cognitive picture and existing practice, we adopt a \(100\)\,ms threshold and ask: on real ten-finger typing data, how often do adjacent \emph{intentional} keystrokes fall within this window? Our goal is to verify that treating \(\leq 100\)\,ms as near-synchronous merges only a small fraction of genuine successive keystrokes.}

\tl{We re-analyze the user dataset from \emph{ResType} \cite{li2023restype}, focusing on Study~2. In that study, participants (8 users) typed MacKenzie phrases \cite{mackenzie2003phrase} on a fixed QWERTY layout rendered on a Sensel Morph touchpad \cite{jacobi2019sensel}, while the same unintentional-touch prevention as \textit{TypeBoard} \cite{gu2021typeboard} filtered resting contacts. The log provides per-event timestamps, contact states, and intentionality labels at a fixed sampling rate of \(50\)\,fps. We use these labels and timestamps only, as this condition best matches our scenario of ten-finger typing on a standard soft keyboard. Figure~\ref{fig:touch-heatmaps} shows the spatial distributions: intentional touches cluster over keys, whereas unintentional contacts form a smooth resting arc.}


\begin{figure}[t]
  \centering
  \includegraphics[width=\linewidth]{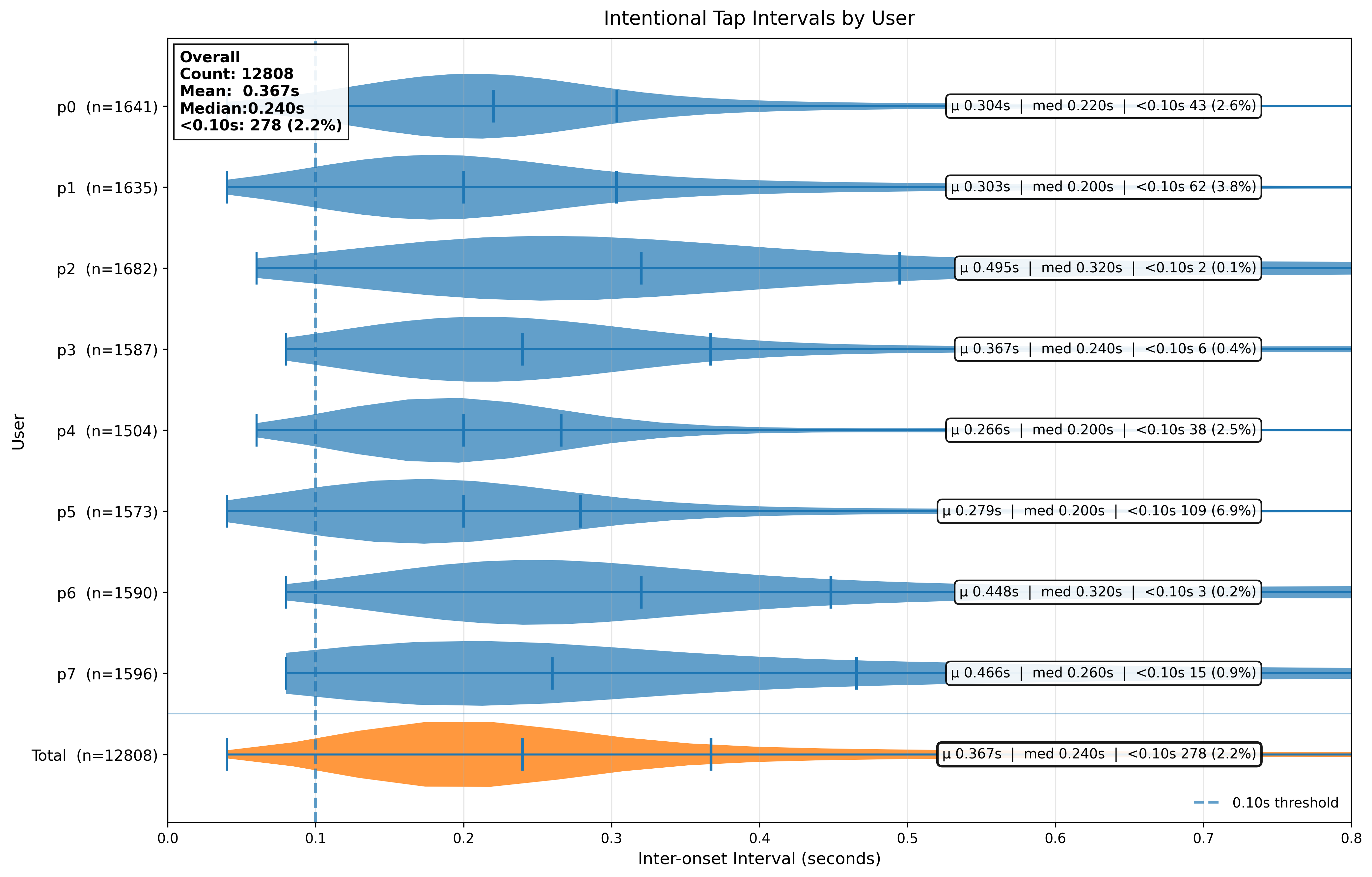}
  \caption{Inter-onset intervals by user (violins) with a 100\,ms reference line. The rightmost violin aggregates all users. Labels show counts and percentages of gaps $\leq 100$\,ms. Means (green) and medians (red) are overlaid.}
  \label{fig:interval-violin}
  \Description{Inter-onset intervals of intentional keystrokes by user with a 100 ms reference line. Violin plots show very few adjacent keystrokes ≤100 ms (≈2\%), supporting a 100 ms clustering threshold; medians are around a few hundred milliseconds.}
\end{figure}

\tl{For each participant, we extract the onset time of every intentional contact and compute the adjacent inter-onset gaps \(\{\Delta t_m\}\). Our primary statistic is the fraction of gaps at or below \(100\)\,ms, i.e., the proportion of successive taps that would be merged if \(\leq 100\)\,ms were treated as synchronous. Across \(U=8\) participants and \(M=12{,}808\) adjacent intentional gaps, only \(\mathbf{2.17\%}\) of intervals are \(\leq 100\)\,ms. A user-stratified bootstrap yields a 95\% confidence interval of \(\mathbf{[0.87\%,\,3.75\%]}\), indicating that this mass is consistently small across users. Figure~\ref{fig:interval-violin} visualizes the per-user gap distributions and highlights the small share below 100\,ms; the pooled median inter-key interval is about \(\mathbf{240}\)\,ms, with most gaps falling well above this threshold. A sensitivity check that jitters the 100\,ms cutoff by \(\pm 10\)\,ms produces nearly identical percentages, suggesting that 20\,ms frame quantization has negligible effect on this conclusion.}

For system design, these numbers mean that adopting \(\tau_c=100\)\,ms as a merging window will fuse only about \(2\%\) of adjacent intentional keystroke pairs. We therefore use \(\tau_c=100\)\,ms as a principled, user-agnostic default in our pipeline. 
\subsection{Thread Formation and Time Clustering}
\label{subsec:thread-cluster}

This subsection introduces the first two stages: (i) reconstruct coherent \emph{touch threads} from the raw event stream, then (ii) group near-synchronous threads into short-lived time clusters.

\paragraph*{Thread formation from the event stream.}
The logger yields a sequence of low-level events \(e_k=(\texttt{type}_k,t_k,\mathbf{z}_k)\), where \(\texttt{type}\in\{\texttt{down},\texttt{move},\texttt{up}\}\), \(t_k\) is the timestamp, and \(\mathbf{z}_k=(x_k,y_k)\) is a normalized location. We maintain a set of open threads, each storing its first/last timestamps and endpoints and a polyline of intermediate moves. \tl{A \texttt{down} event starts a new thread; subsequent \texttt{move} and \texttt{up} events are associated to the nearest open thread whose endpoint is within a small spatial and temporal neighborhood, and close the thread when an \texttt{up} is seen.}

We perform data association via a light-weight spatial hash.
The keyboard plane is partitioned into a uniform grid of cell size \(h\) and index the current endpoints of all open threads.
Given an event at \(\mathbf{z}_k\), we probe the occupied entries in the cell containing \(\mathbf{z}_k\) and its 8 neighbors, and select the nearest open thread whose last endpoint within a spatial radius \(r_g\) and a short
temporal gap \(T_{\text{gap}}\).
If found, we append \(\mathbf{z}_k\) to that thread and update its last endpoint; otherwise, we treat the event as a stray contact (ignored) or the start of a new thread.
On a matched \texttt{up}, we \emph{close} the thread and emit
\(\mathbf{t}_i=\big(t_i^{\text{start}},t_i^{\text{end}},\mathbf{x}_i^{\text{start}},\mathbf{x}_i^{\text{end}},\text{id}_i\big)\),
where \(\mathbf{x}_i^{\text{end}}\) is the last endpoint and \(\text{id}_i\) denotes our \emph{internal} thread identifier.

Although iPadOS exposes hardware touch identifiers, in practice they are often unreliable. We therefore treat all events as \emph{anonymous} touches and perform aggregation via the spatial hash, which yields \(O(1)\) expected-time association per event \tl{and produces coherent threads in real time.}

\paragraph*{Time clustering.}
\tl{Given the set of threads for a word, we sort threads by onset time and grow clusters greedily. Intuitively, a thread joins the current cluster if it starts within the cognitive window \(\tau_c\) of the cluster’s anchor. Large, near-simultaneous multi-finger contacts likely correspond to calibration-like gestures (e.g. place the fingers on the home-row before typing); clusters whose size exceeds a cap \(\kappa\) are discarded.}


\subsection{In-Cluster Selection Using Travel Score}
\label{subsec:selection-travel}

Within each cluster, we select the representative touch of user's intent. Our heuristic is informed by motor control principles: we favor deliberate \emph{reaches}, spatially significant departures over incidental, low-motion contacts.



\tl{We model the hand's recent state as a \emph{hand-state cloud} \(\mathcal{H}(t)\),
a decaying memory of recent finger endpoints. Each point \(\mathbf{p}_m\) in the
cloud stores its location and time stamp. Points older than a horizon
\(T_{\max}\) are pruned. The influence of each prior point decays
geometrically over time with factor \(\rho\) and step \(\Delta\), so more
recent contacts carry higher weight.}

For each candidate touch \(i\) in a cluster, with endpoint \(\mathbf{x}_i\) and
onset time \(t_i\), we compute a \emph{travel score} measuring its spatial
departure from the cloud, down-weighted by the age of the nearest prior contact:
\begin{equation}
S_i=\min_{(\mathbf{p}_m,\tau_m)\in \mathcal{H}(t_i)}
\frac{\lVert \mathbf{x}_i - \mathbf{p}_m\rVert_2}{\max\{w_m(t_i),\epsilon\}},
\label{eq:travel}
\end{equation}
where \(w_m(t_i)\) is the decayed weight of point \(m\) at time \(t_i\) and
\(\epsilon\) is a small constant to avoid division by zero. Larger scores
correspond to more distinct and more recent reaches. Conceptually, this score
represents a time-discounted distance: large, fresh movements away from the
recent pattern are scored highest.
We then choose the intended activation as:
\begin{equation}
    i^*=\arg\max_{i\in C} S_i.
\end{equation}
To handle cases where travel scores are nearly tied (e.g., fast in-place taps), we select the most recent touch in the cluster. \tl{This rule leverages a characteristic pattern of ten-finger posture: when the user reaches toward a target key, nearby resting fingers are often slightly \emph{dragged} in the same direction and tend to land earlier with shorter travel, producing premature incidental contacts, whereas the true intentional press occurs last. The same logic also governs stationary taps on home-row keys, which exhibit minimal travel but still appear as the final contact in their cluster.}

\begin{figure*}[t]
 \centering
 \includegraphics[width=\linewidth]{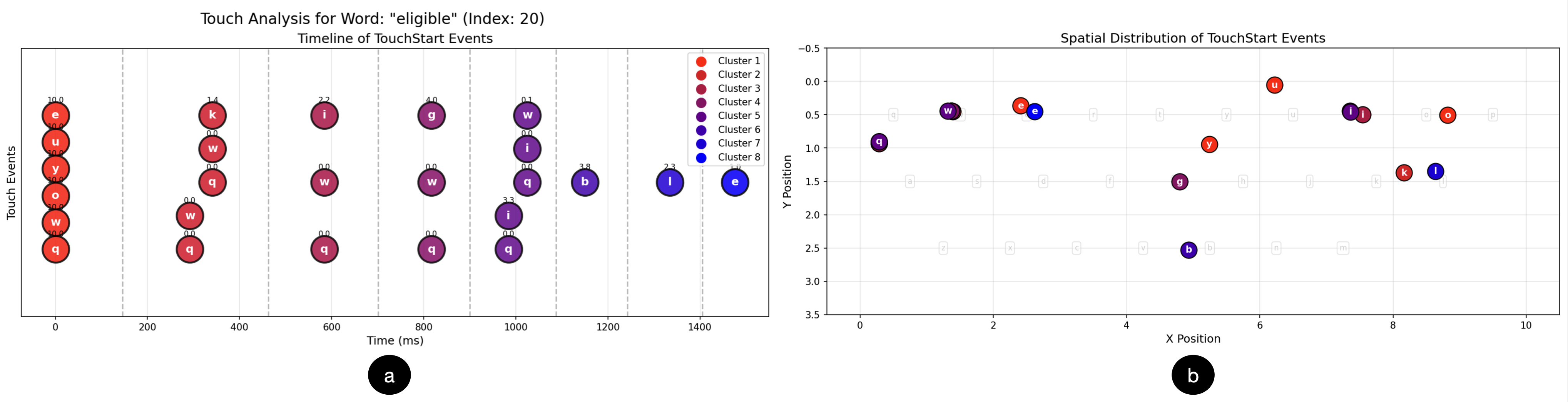}
 \caption{Cluster-level selection for a representative word instance (\emph{eligible}). \textbf{(a)} timeline with cluster assignments; \textbf{(b)} spatial distribution of cluster start positions with nearest keys faintly shown. Our in-cluster rule yields \texttt{ekigible}.}
 \label{fig:cluster-example}
 \Description{Example selection for the word “eligible.” Timeline and spatial panels show eight time-compact clusters; the heuristic yields the intermediate “ekigible” (k for l), a near-neighbor substitution consistent with small lateral drift—the kind of noise the decoder fixes. The example illustrates how near-synchronous touches are merged into clusters.}
\end{figure*}

\paragraph*{Example.}
Figure~\ref{fig:cluster-example} illustrates the process for the word \emph{eligible}. Eight clusters are formed, and their selected representatives yield the intermediate sequence \texttt{ekigible}. The substitution (\texttt{k} for \texttt{l}) arises from a small lateral drift, which is one type of structured noise addressed by our downstream decoder component.

\paragraph*{Implementation.}
The parameters (e.g. cognitive window $\tau_c$, hand-state cloud’s decay parameters ($\rho,\Delta, T_{\max}$) were optimized via a simple grid search. The resulting pre-decoding stage produces a letter sequence with structured, local noise typical of this interaction style: adjacent-key substitutions and merge/split artifacts from temporal clustering.

\section{Synthetic Noise Modeling and LLM Fine-Tuning}

With front-end producing letter sequences, we now address mapping the noisy letter sequence to a word. We first collect touch logs of hands-down typing behavior, then fit a simple noise model that generates synthetic word pairs, and finally fine-tune a LLM (FLAN--T5-small) for correction.

\subsection{Data Collection for Behavior Modeling}
\label{subsec:collection}

We collected touch logs as participants typed prompted words on a tablet. The goal was to capture event-level traces (touch start/move/end timestamps and locations) under realistic ten-finger posture so we could characterize where near-synchronous and reach-induced incidental contacts occur.

\begin{figure*}[t]
  \centering
  \includegraphics[width=0.8\linewidth]{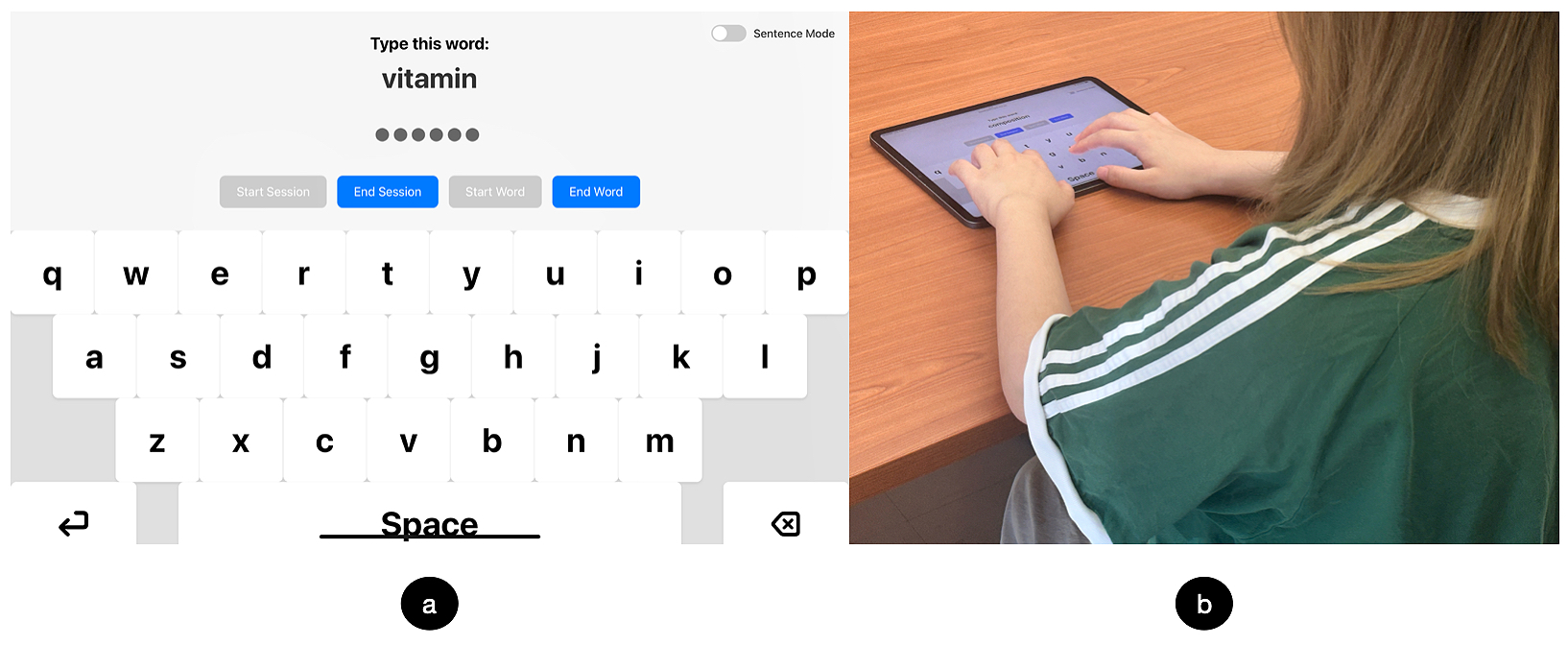}
\caption{Word collection on an iPad Pro. 
(a) A target word appears at the top; instrumentation gates precise logging. \tl{During typing, the interface shows only a password-like progress bar (gray dots) instead of intermediate letters, encouraging natural input without mid-word corrections.}
(b) A participant types with all ten fingers resting on the surface.}
  \label{fig:word-collect}
  \Description{Data-collection interface on iPad Pro with obfuscated in-progress feedback. (a) The target word appears at the top; the keyboard logs start/move/end events. A password-like row of dots indicates progress instead of raw letters to avoid distraction and premature self-correction. (b) A participant demonstrates the typing posture, resting all fingers on the tablet while entering the target word.}
\end{figure*}

\paragraph*{Apparatus and Participants.}
We implemented a web-based typing interface on an 11-inch iPad Pro \cite{apple_ipad_pro_11} using a standard QWERTY layout (Figure~\ref{fig:word-collect}a). The logger records all touch lifecycle events with millisecond timestamps and normalized coordinates. Eight volunteers (7 male, 1 female), aged 21--26 years (\textit{M}~=~23.38, \textit{SD}~=~1.60), participated with informed consent. They were instructed to rest all ten fingers on the surface while typing (Figure~\ref{fig:word-collect}b).

\paragraph*{Procedure}
Each session comprised blocks of \emph{word} prompts sampled from google-10000-english \cite{google10000english_dataset}. For each prompt, participants tapped \textsf{Start}, typed on the keyboard, and tapped \textsf{End}. Avoiding mid-word corrections allowed the logger to capture incidental and near-synchronous contacts that the decoder would later resolve.

\paragraph*{Obfuscated in-progress feedback.}
To prevent participants from reacting to noisy intermediate sequences, the interface presented only password-like progress dots (Figure~\ref{fig:word-collect}a). This offered minimal awareness without revealing errors caused by overlapping touches or reach dynamics, preserves natural behavior \cite{bi2013octopus}. \tl{The obfuscated feedback was used only during data collection; all later user studies used visible intermediate feedback (Section~\ref{subsec:exp-procedure}).}

\paragraph*{Dataset.}
We collected \textbf{1{,}752} prompted words. Word lengths were well distributed: $\leq\!3$ letters 8.7\%, 4--6 letters 41.4\%, 7--9 letters 37.2\%, and $\geq\!10$ letters 12.8\%. For every contact we logged the event type, timestamp, normalized coordinates, and session/word identifiers. This dataset covers both short bursts (where near-synchronous touches are more common) and longer words with richer hand travel, \tl{and serves as the basis for parameterizing our synthetic error generator.}

\subsection{Synthetic Training Data Generation}
\label{subsec:synth}

\tl{Real-world error data from hands-down typing are sparse and unevenly distributed. We therefore augment them with synthetic noisy--clean pairs \cite{sennrich2015improving,xie2017data,wei2019eda} that mimic the physical structure of our interaction while spanning a controlled range of difficulties. Each pair consists of a target English word \(w\) (sampled from \texttt{google-10000-english} \cite{google10000english_dataset}) and a noisy letter sequence \(u\) that simulates what our pre-decoder would output for that word. We write \(\mathrm{ED}(u,w)\) for their Levenshtein edit distance.}


\paragraph*{From human pairs to a simple error model.}
We start from human noisy--clean pairs collected in our study and align them with Levenshtein distance to extract atomic edits (substitutions, insertions, deletions, and local swaps). For each position we attach simple geometry features from the QWERTY layout (row, hand, finger assignment, local travel distance). Positions that require long reaches or unstable postures thus receive higher edit propensity.

\paragraph*{Two dominant error channels.}
Empirically, hands-down typing errors are dominated by two mechanisms: near-key slips of the intended finger and co-activation of nearby resting fingers.

\emph{Near-key slips (\textsc{Near}).}
\tl{When the user aims at a key for letter \(a\), the active finger may land slightly off its center. For each letter \(a\), we fit a small 2D Gaussian mixture model (GMM) over fingertip landing offsets \(\boldsymbol{\delta}\), obtained from human touch data. Let \(\mathbf{p}(a)\) denote the center of key \(a\) in keyboard coordinates, and let the perturbed contact location be
\begin{equation}
    \tilde{\mathbf{p}}=\mathbf{p}(a)+\boldsymbol{\delta}.
\end{equation}
We then map \(\tilde{\mathbf{p}}\) back to letters with a soft nearest-key kernel
\begin{equation}
q(c\mid\tilde{\mathbf{p}})\propto \exp\!\ \bigl(-\alpha\|\tilde{\mathbf{p}}-\mathbf{p}(c)\|^2\bigr),
\label{eq:soft-key}
\end{equation}
where \(\alpha>0\) controls how sharply probability decays with spatial distance. Taking the most probable letter under \(q\) yields plausible \textbf{substitutions}, sampling from \(q\) yields occasional \textbf{extra taps} around the target key, and a special “no-key’’ outcome models \textbf{deletions} when the slip falls between keys. In practice, a small number of mixture components per letter suffices to capture typical offset patterns.}

\emph{Co-activation of resting fingers (\textsc{CoAct}).}
\tl{When one finger moves, other fingers resting nearby may briefly touch the surface. To capture this, we examine clusters of concurrent touches in the human data and, for each target letter \(a\) and coarse geometry bucket (e.g., row\(\times\)hand), count how often other touches within the same cluster fire as letter \(c\). From these counts we fit a smoothed categorical distribution \(P_{\mathrm{CoAct}}(c\mid a,\text{bucket})\), which describes, for a given intended letter and posture, which other letters are likely to appear as co-activated taps. During synthesis, sampling from this distribution inserts extra letters (and can replace the aimed letter with small probability), and we allow occasional local swaps between adjacent positions, with probabilities that decrease as cross-row or cross-hand distance grows. This channel introduces realistic multi-letter bursts and small re-orderings that pure geometry cannot explain.}

\paragraph*{Synthesis.}
Given a target word \(y\) of length \(L\), we generate a synthetic noisy sequence \(x\) in three steps.

\emph{(1) Choose how many edits to apply.}
We draw an intended edit count from length-aware prior and cap difficulty by
\begin{equation}
E_{\max}(L)=\begin{cases}
2,&L\le6,\\
3,&7\le L\le9,\\
4,&L\ge10,
\end{cases}
\label{eq:ed-cap}
\end{equation}
so that very short words are not overwhelmed by edits, while longer words can exhibit richer corruptions.

\emph{(2) Select positions and instantiate errors.}
\tl{Positions in \(y\) are sampled according to their learned edit propensities.} For each selected position we choose an error channel (\textsc{Near} or \textsc{CoAct}) based on coarse geometry/context, assign an operation type (substitution, insertion, deletion, or swap), and instantiate its content using the corresponding mechanism: Near-key slips generate letters via the soft kernel in \eqref{eq:soft-key}, and co-activation events draw from the empirical \(P_{\mathrm{CoAct}}\). Operations are applied in a fixed, simple order (swaps, then deletions, then insertions/substitutions from left to right) to obtain a candidate noisy string \(x\). Because operations can cancel or interact, the realized edit distance \(\mathrm{ED}(x,y)\) may differ slightly from the initial intended count.

\emph{(3) Balance difficulty across edit distances.}
To avoid overproducing trivial near-copies and to expose the model to a range of difficulties, we target an approximately uniform distribution over edit distances up to \(E_{\max}(L)\):
\begin{equation}
\widehat{\Pr}\!\big(\mathrm{ED}(x,y)=e\mid|y|=L\big)\approx\frac{1}{E_{\max}(L)+1},
\label{eq:balanced-ed}
\end{equation}
and accept only samples with \(\mathrm{ED}(x,y)\le E_{\max}(L)\). We resample until the empirical edit-distance histogram for each length regime is close to the target. Mild randomness in the GMM offsets, co-activation counts, and sampling steps injects diversity while keeping the overall statistics aligned with human data.

This process yields \(\mathbf{146{,}331}\) synthetic pairs over \(\mathbf{7{,}700}\) unique target words. By explicitly modeling both geometry-driven slips and concurrent touches, and by balancing edit distance across length regimes, the generator presents the LLM with a realistic and difficulty-controlled corruption manifold. 

\subsection{Model Fine-Tuning}
\label{subsec:fine-tune}

We cast single-word correction as conditional sequence generation: given a corrupted character sequence \(x\) from our synthetic generator (Sec.~\ref{subsec:synth}), an fine-tuned LLM \(p_{\theta}\) (FLAN--T5-small \cite{wei2021finetuned,chung2022scaling}, 77M parameters) predicts the intended word \(y\). \tl{Although the noise-producing model operates at the character level, FLAN--T5 internally processes \emph{subword tokens} produced by its SentencePiece tokenizer. We lowercase each noisy string and tokenize it into a short subword sequence (e.g., ``eligible'' \(\rightarrow\) [``eli'', ``gi'', ``ble'']), and train the decoder to map these tokenized inputs back to a single word. Throughout this paper we use “characters’’ for elements of \(x\) and “tokens’’ for FLAN subwords.}

\paragraph*{Model choice.}
\tl{We adopt FLAN--T5-small because its encoder--decoder Seq2Seq design matches our ``noisy sequence to one word'' objective: the encoder produces a fixed representation of the corrupted input, and the decoder attends over it to generate the corrected word, reducing copy bias and instruction leakage. The small 77M-parameter model trains quickly on a single GPU and supports real-time inference for interactive decoding.}

\paragraph*{Training objective.}
Each training pair maps one noisy sequence to one lowercase English target word. We fine-tune all layers end-to-end under teacher forcing and minimize the negative log-likelihood
\begin{equation}
\mathcal{L}_{\mathrm{NLL}}(\theta)
= - \sum_{i=1}^{N} \log p_{\theta}\!\big(y_i \,\big|\, p(x_i)\big),
\label{eq:nll}
\end{equation}
where \(p(x_i)\) denotes the tokenized input. To gently bias the model toward orthographically close outputs, we add a small surrogate penalty on the edit distance between the teacher-forced prediction \(\widehat{y}_i\) and the target:
\begin{equation}
\mathcal{L}_{\mathrm{ED}}(\theta)
= \lambda_{\mathrm{ED}} \cdot \frac{1}{N}\sum_{i=1}^{N} \mathrm{ED}\!\big(\widehat{y}_i, y_i\big),
\label{eq:ed-loss}
\end{equation}
and optimize \(\mathcal{L}=\mathcal{L}_{\mathrm{NLL}}+\mathcal{L}_{\mathrm{ED}}\). In practice, most gains come from the synthetic curriculum so \(\lambda_{\mathrm{ED}}\) is kept small. \tl{We prepend a short spell-correction prompt during training to preserve the instruction-following context of FLAN models, but performance is stable across different prefixes, indicating that the model learns primarily from the synthetic error distribution rather than from the exact wording of the instruction.}

\paragraph*{Optimization and implementation.}
We fine-tune using AdamW \cite{loshchilov2017decoupled} with a linear learning-rate schedule and 5\% warmup: learning rate \(5{\times}10^{-4}\), weight decay \(0.01\), batch size \(32\) for training and \(64\) for validation. Training runs for 16~epochs with a 95/5 train/validation split. The pipeline is implemented in PyTorch and Hugging Face Transformers \cite{paszke2019pytorch,wolf2020transformers} and trained on a single 48\,GB NVIDIA RTX~A6000 GPU.

\paragraph*{Validation metrics.}
For validation we report \emph{exact match} (EM),
\begin{equation}
\mathrm{EM}
=\frac{1}{M}\sum_{i=1}^{M}\mathbf{1}\{\widehat{y}_i=y_i\},
\label{eq:eval-metrics}
\end{equation}
and additionally monitor the single-word validity rate to detect multi-token or non-alphabetic outputs. \tl{Because character-level typos can induce large changes in the model’s subword sequence, the LLM effectively learns to repair tokenization-distorted inputs rather than apply simple character edits. This also clarifies our following comparison with the character-based Bayesian baselines.}
\section{Understanding Decoding Performance on Synthetic Data}
\label{sec:synthetic-decoding}

We evaluate the LLM decoding component on a synthetic corpus of misspelled–gold pairs (Sec.~4.1). We disable cross-word context, so the comparison isolates phrase-level hints, matching prior statistical decoders \cite{weir2014uncertain,vertanen2015velocitap}.

\subsection{\xj{Two Baseline Decoders}}
\label{subsec:bayes-baseline}

\tl{
We compare our LLM decoder against two baselines.  
The primary baseline is a \xj{touch-location informed Bayesian decoder}, the standard text entry decoder that leverages full touch geometry~\cite{goodman2002language, zhu2018typing}.  
To match the same letter-only input consumed by LLM, we also include a \xj{letter-only n-gram decoder} that operates solely on the nearest-key sequence.}



\xj{
\subsubsection*{Touch-location Informed Bayesian Decoder}
We implemented this decoder following the commonly used Bayesian decoding principle (a.k.a, statistical decoding principle)~\cite{goodman2002language, zhu2018typing}.
More specifically, given a sequence of touch points $S = \left\{s_1, s_2, s_3, ..., s_n\right\}$, this decoder is to find word $w^*$ in lexicon $L$ that satisfies:
\begin{equation}\label{eq:decoder}
    w^* = \underset{w \in L}{\arg\max}\ P(w|S).
\end{equation}

Using the Bayesian rule and assuming $P(S)$ is an invariant across words, this equation becomes:
\begin{equation}\label{eq:decoder-final}
    w^* = \underset{w \in L}{\arg\max}\ \frac{P(S|w)P(w)}{P(S)} = 
    \underset{w \in L}{\arg\max}\ P(S|w)P(w),
\end{equation}
where $P(w)$ is from a language model (LM) and $P(S|w)$ is from a spatial model (SM). Similar to common practice (e.g., \cite{zhu2018typing}), We adopted a bivariate Gaussian touch model as our spatial model, and used a character \(n\)-gram model ($n=5$) with standard add-\(k\) smoothing as our language model.} \tl{Here, the LM serves as a lightweight lexical prior over candidate words, while the dominant contribution comes from spatial likelihoods and alignment constraints; larger \(n\) yields diminishing returns under this edit-bounded setting.}

\subsubsection*{Letter-only N-gram Decoder}




\tl{When spatial coordinates $S$ are not available, this decoder selects word $w$ using the nearest-key sequence $u$ from candidate set $C(u)=\{\,w\in D:\ \mathrm{ED}(u,w)\le 4\,\}$. $\mathrm{ED}(\cdot,\cdot)$ denotes the editing distance. Following equations \ref{eq:decoder} and \ref{eq:decoder-final} and treating $P(u)$ as constant, the optimal word is:

\begin{equation}
w^* = 
\underset{w \in C(u)}{\arg\max}\  P(u \mid w)P(w)
=
\underset{w \in C(u)}{\arg\max}\  [\log P(u\mid w) + \log P(w)].
\label{eq:argmax-u}
\end{equation}

In the absence of spatial likelihoods, $P(u \mid w)$ rely solely on how well the nearest-key sequence matches a candidate word. We approximate $\log P(u \mid w)$ using the negative editing distance between $u$ and $w$:
\begin{equation}
w^*
=
\underset{w \in C(u)}{\arg\max}\  [\log P(w)
- \alpha \cdot \mathrm{ED}(u,w)],
\label{eq:likelihood-ed}
\end{equation}
where $\alpha$ is a control parameter that balances language prior and matching likelihood, which has been tuned.}

\begin{figure}[t]
  \centering
  \includegraphics[width=\linewidth]{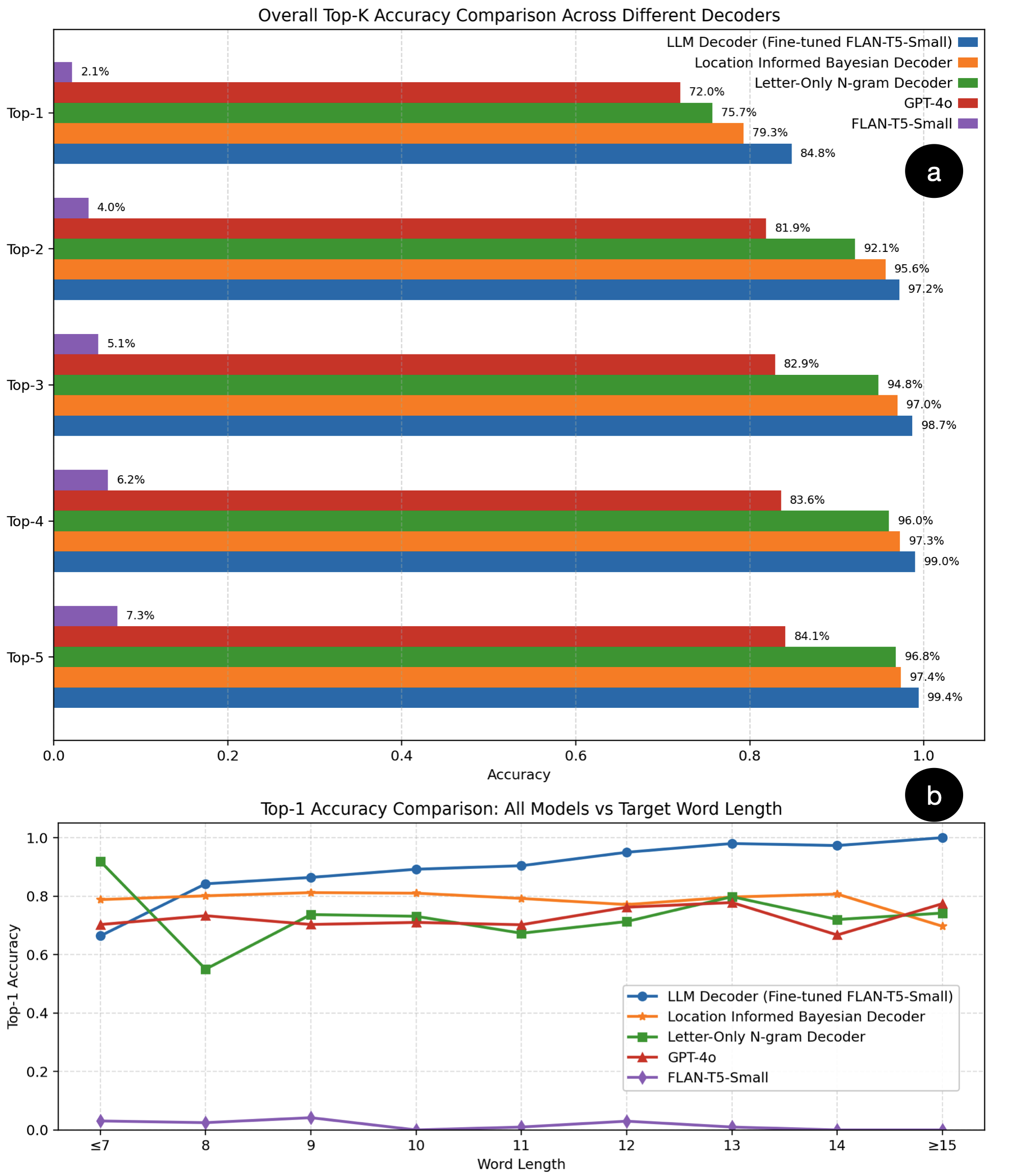}
  \caption{\textbf{(a)} Overall Top-$k$ accuracy comparison across five decoders.  
The Spatial Bayesian variant narrows the gap between the standard statistical model and the fine-tuned LLM decoder by 3–4 points at Top-1. The LLM still achieves the best ranking and lowest residual error. \textbf{(b)} Top-1 accuracy versus word length shows that the spatial model helps most for short words, while the LLM retains superiority for medium and long words.}

  \label{fig:overall-topk}
  \Description{Decoding accuracy across systems and lengths. Bars and lines summarize Top-k accuracy across decoders and word lengths, showing the fine-tuned FLAN-T5-small achieving the highest overall Top-1 accuracy (≈84.8\%) and near-ceiling Top-5, outperforming a Bayesian baseline and GPT-4o; the largest gains appear on medium/long words.}
\end{figure}

\subsection{Decoding Accuracy Comparison}
\label{subsec:decoding-accuracy}

\paragraph{Evaluation setup.}
We evaluate on the synthetic misspelling–gold pairs (Sec.~\ref{subsec:synth}).  
The touch-location informed Bayesian Decoder and the letter-only n-gram Decoder (Sec.~\ref{subsec:bayes-baseline}) are chosen as baselines.  
For LLMs, candidate scoring is obtained via decoding with beam width $k$; beam search exposes the highest-probability $k$ hypotheses. 
\tl{This multi-hypothesis retrieval requires no additional fine-tuning.}

\paragraph{Overall accuracy.}
Figure~\ref{fig:overall-topk}\,a summarizes aggregate performance across all decoders.  
The fine-tuned LLM decoder attains \textbf{84.8\%} Top-1 and \textbf{97.2/98.7/99.4\%} at Top-2/3/5.  
The letter-only n-gram baseline reaches \textbf{75.7\%} Top-1 and \textbf{92.1/94.8/96.8\%}.
\tl{Adding touch-location improves the Bayesian decoder to \textbf{79.3\%} Top-1 and \textbf{95.6/97.0/97.4\%}. Yet the LLM, which receives only the letter sequence, still ranks candidates more accurately, showing that its advantage stems from stronger sequence-level priors rather than spatial cues.}

\begin{figure*}[t]
  \centering
  \includegraphics[width=0.95\linewidth]{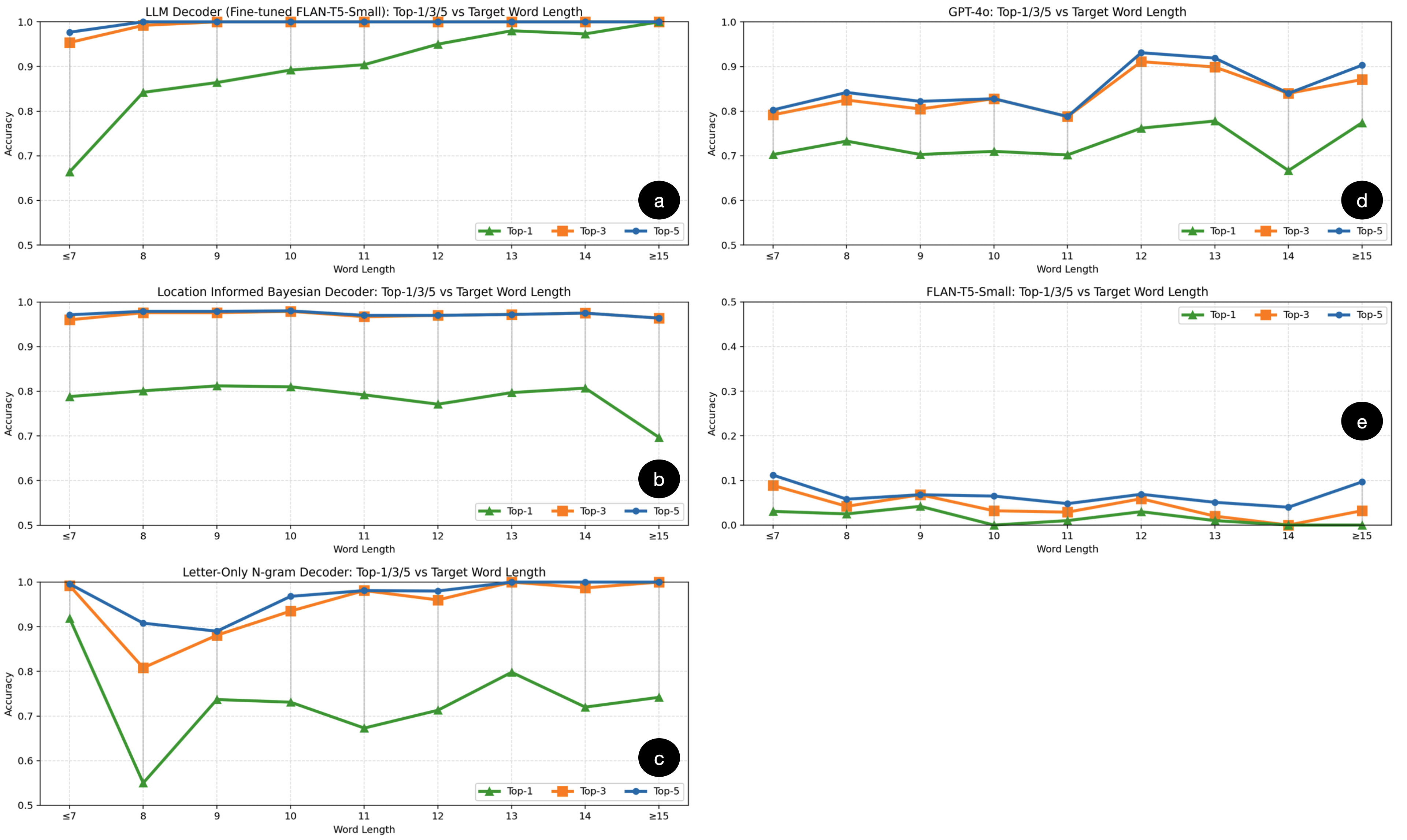}
  \caption{\textbf{Top-1/3/5 accuracy versus target length for each decoder (a–e).}  
(a) Fine-tuned FLAN–T5 (LLM decoder) quickly closes the Top-1 and Top-5 gap as length increases.  
(b) Bayesian decoder benefits short words via coordinate-based likelihoods but plateaus earlier on longer words.  
(c) N-gram decoder relies solely on symbolic letter sequences and shows larger length sensitivity.  
(d) GPT-4o remains steady but below Bayesian variants without lexicon constraints.  
(e) Un-fine-tuned FLAN–T5 fails to capture the task semantics.}
  \label{fig:length-topk}
  \Description{Top-1/3/5 accuracy versus word length for each decoder. The fine-tuned LLM’s Top-1–Top-5 gap shrinks as length grows, indicating sharper ranking; the Bayesian baseline is strongest only on short words.}
\end{figure*}

\paragraph{Effects of word length.}
Figure~\ref{fig:length-topk} details the dependence on target length; for reporting we aggregate to three buckets (\(\leq\!7\), \(8\!-\!14\), \(\geq\!15\); cf.\ Sec.~\ref{sec:synthetic-decoding}).  
\emph{Short (\(\le7\) chars).}
Local regularities dominate and are well captured by a character $n$-gram prior; the n-gram baseline is strongest (Top-1 \(\approx\)\,\textbf{91.9\%}), while our LLM trails (Top-1 \(\approx\)\,\textbf{66.4\%}).  
\tl{The Bayesian decoder reaches \textbf{80.0\%} Top-1. Short words are already well separated by symbolic patterns alone, so injecting noisy coordinate likelihoods offers limited benefit.}

\emph{Medium (\(8\!-\!14\) chars).}
Ambiguity grows as the candidate set expands with length \(L\); windowed $n$-grams underfit longer-range orthography and morphology.  
Here the LLM reaches \textbf{91.0\%} Top-1 vs.\ \textbf{69.9\%} for the n-gram baseline, and the Top-1–Top-5 gap narrows sharply.  
\tl{Spatial cues offer marginal help but cannot compensate for the absence of global linguistic priors, leaving the Spatial Bayesian decoder far below the LLM at these lengths.}

\emph{Long (\(\ge15\) chars).}
With more global cues (e.g., \emph{-tion}, \emph{-ment}), the LLM’s sequence prior excels, achieving \textbf{100\%} Top-1 while the n-gram baseline yields \textbf{74.2\%}.  
\tl{The Bayesian decoder drops to \textbf{68.6\%} Top-1, reflecting the difficulty of aligning long, dispersed touches.}  
Overall, spatial likelihoods help moderate-length words but fade on very short or very long ones, while the LLM maintains robust performance across all lengths.

\paragraph{Why the fine-tuned LLM wins.}
\textbf{(i) Sequence-level priors.} \tl{Fine-tuning gives the model global orthographic and morphological priors.
Statistical decoders rely on local symbolic or geometric cues, leading to $n$-gram underfitting. 
LLM errors, in contrast, tend to be morphology-preserving near misses.}
\textbf{(ii) Task-shaped output space.} \tl{Fine-tuning aligns the model’s SentencePiece segmentation with our one-word objective, suppressing multi-token drift and producing stable outputs.}
\textbf{(iii) Learned conservatism.} Exposure to balanced edit distances (Sec.~\ref{subsec:synth}) encourages small, plausible corrections; even when wrong, outputs remain close (AvgED 0.318).



\paragraph{Specialization vs.\ scale.}
GPT-4o reaches \textbf{72.0\%} Top-1 and \textbf{84.1\%} Top-5, while an un-fine-tuned FLAN--T5-small performs poorly (Top-1: \textbf{2.1\%}, Top-5: \textbf{7.3\%}). GPT-4o’s free-form generation tends to hedge with plausible but lexicon-mismatched guesses, and the small untuned model exhibits copy bias and instruction bleed-through. \tl{Fine-tuning FLAN--T5-small aligns its subword segmentation and teaches robust edit patterns, yielding a compact, task-specialized decoder that decisively outperforms both a much larger general model and its own untuned initialization.}


\paragraph{Takeaways.}
\tl{Across all settings, the fine-tuned FLAN--T5-small achieves the best accuracy, especially on medium and long words, even though it sees only the letter sequence. Spatial likelihoods remain insufficient to close the gap, and a large general model (GPT-4o) underperforms a small task-specialized model on this task. These patterns motivate our choice to deploy the compact fine-tuned LLM decoder component.}

\section{User Study}

\tl{We conducted a controlled within-subjects lab study to evaluate KeySense in realistic transcription tasks and to characterize how people adapt to hands-down, ten-finger surface typing. Our study examined: (1) typing speed, (2) errors and corrections, (3) user behavior and subjective rating, and (4) decoding latency.}

\subsection{Participants and Apparatus}
\label{subsec:participants}

\paragraph*{Participants.}
12 volunteers (10 male, 2 female; age 21–28, \textit{Mean}=24.73, \textit{SD}=2.45) participated, with prior experience in both physical and virtual keyboards (physical: \textit{Mean}=14.25 years, \textit{SD}=3.05; on-screen: \textit{Mean}=6.75 years, \textit{SD}=4.90). All provided informed consent.


\paragraph*{Apparatus.}
The study used a web-based prototype on an 11-inch iPad Pro \cite{apple_ipad_pro_11} via Safari. The interface presented a standard QWERTY layout and logged full touch lifecycles (down/move/up with timestamps and normalized coordinates). LLM decoding ran on an NVIDIA RTX A6000 GPU. No additional sensors were used.

\subsection{Experiment Design and Procedure}
\label{subsec:exp-procedure}

\begin{figure*}[t]
  \centering
  \includegraphics[width=0.8\linewidth]{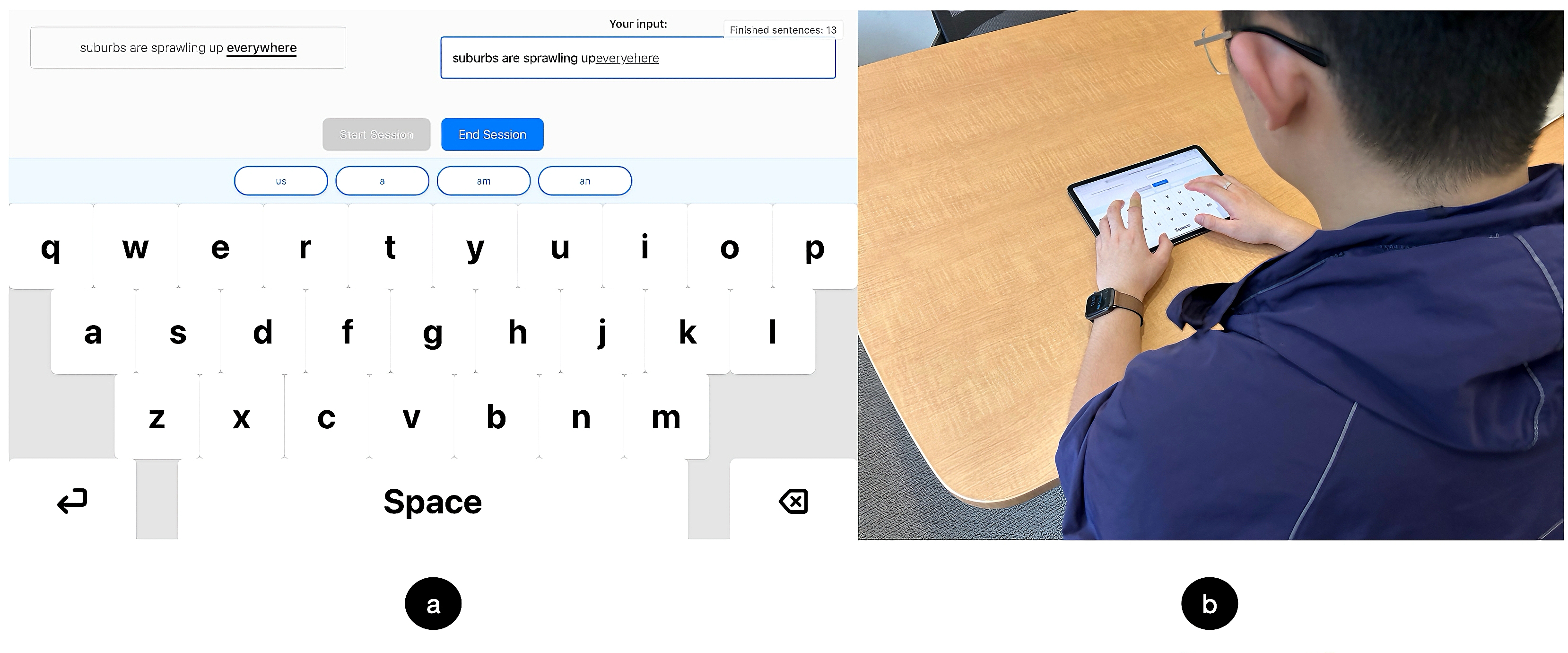}
  \caption{User study setup. 
\textbf{(a)} Transcription interface: the target phrase (left) and the user’s input (right). The intermediate letter sequence is shown live underlined. 
\tl{In this example, the underlined word ``everywhere'' is still in the composing mode in which the decoding has not started yet, and the suggestion bubbles have not been updated and are deactivated.  Once the user presses \textsf{Space}, the system sends the input sequence to the decoder, replaces the underlined segment with the finalized Top-1 word, refreshes and activates the suggestion bubbles with updated suggestions, and adds space to the output.}  
\textbf{(b)} Participant posture with the tablet flat on a desk; most fingers rest on the surface while one finger reaches control keys (Space/Enter/Backspace/suggestions).
}
  \label{fig:user-study}
  \Description{User-study setup—interface and posture. (a) Phrase transcription UI shows the live intermediate sequence for the current word and candidate suggestions, committing Top-1 on Space. (b) Tablet lies flat; most fingers rest on the surface while one finger reaches for controls.}
\end{figure*}

\paragraph*{Task and sessions.}
We used a within-subjects design with two conditions: (i) a conventional soft-keyboard baseline (hover typing; one contact per key) using a n-gram decoder, and (ii) a ten-finger condition decoded by KeySense. In each condition, participants performed a transcription task with phrases from the MacKenzie and Soukoreff's dataset \cite{mackenzie2003phrase}. Each part contained one warm-up block (10 phrases) followed by five formal sessions (10 phrases each). All participants typed the same phrases; only the order was randomized per participant and condition (Figure~\ref{fig:user-study}a).

\paragraph*{Interface and feedback.}
The interface displayed the target phrase above the user’s input (Figure~\ref{fig:user-study}a). We showed the underlined intermediate letter sequence produced by our pre-decoding pipeline so participants could monitor how their touches were interpreted. \tl{Decoding was triggered only when the user pressed \textsf{Space}: the underlined intermediate sequence was replaced by the Top-1 decoded word, and ranks 2–5 appeared as tap-selectable suggestions above the keyboard (generated via beam search with width 5). This setup receives decoding only after word completion, reducing inference cost and preserving continuous typing flow.}

\paragraph*{Editing controls.}
\textsf{Enter} submitted the current phrase and loaded the next one.  
\textsf{Backspace} acted in two modes: it cleared the current letter sequence if mid-word, or removed the previously committed word otherwise. Participants could rest non-reaching fingers on the surface when controlling as well: only the reaching finger moved to \textsf{Space}, \textsf{Enter}, \textsf{Backspace}, or suggestions.

\paragraph*{Procedure.}
After a brief demonstration, participants first completed one of the two conditions (order counterbalanced), then the other. In the conventional part they hovered and tapped keys individually; in the ten-finger part they were encouraged to rest non-reaching fingers (Figure~\ref{fig:user-study}b). The editing controls were identical across conditions. Participants were instructed to transcribe as \emph{quickly and accurately} as possible. Each condition started with a 10-phrase warm-up and then five formal sessions (50 phrases). After both conditions, participants completed a usability questionnaire.

\subsection{Results}
\label{subsec:user-results}

\subsubsection{Typing speed.}
We report speed in Words-Per-Minute (WPM) following standard practice \cite{bi2012bimanual}:
\begin{equation}
\mathrm{WPM} = \frac{|S_{transcribed}|}{5\times T_{minutes}},
\end{equation}
where $|S_{transcribed}|$ is the length of the final transcription in characters and $T$ is the total time in minutes. Timing begins at the start signal, so the character count is used directly.

\begin{figure}[t]
  \centering
  \includegraphics[width=\linewidth]{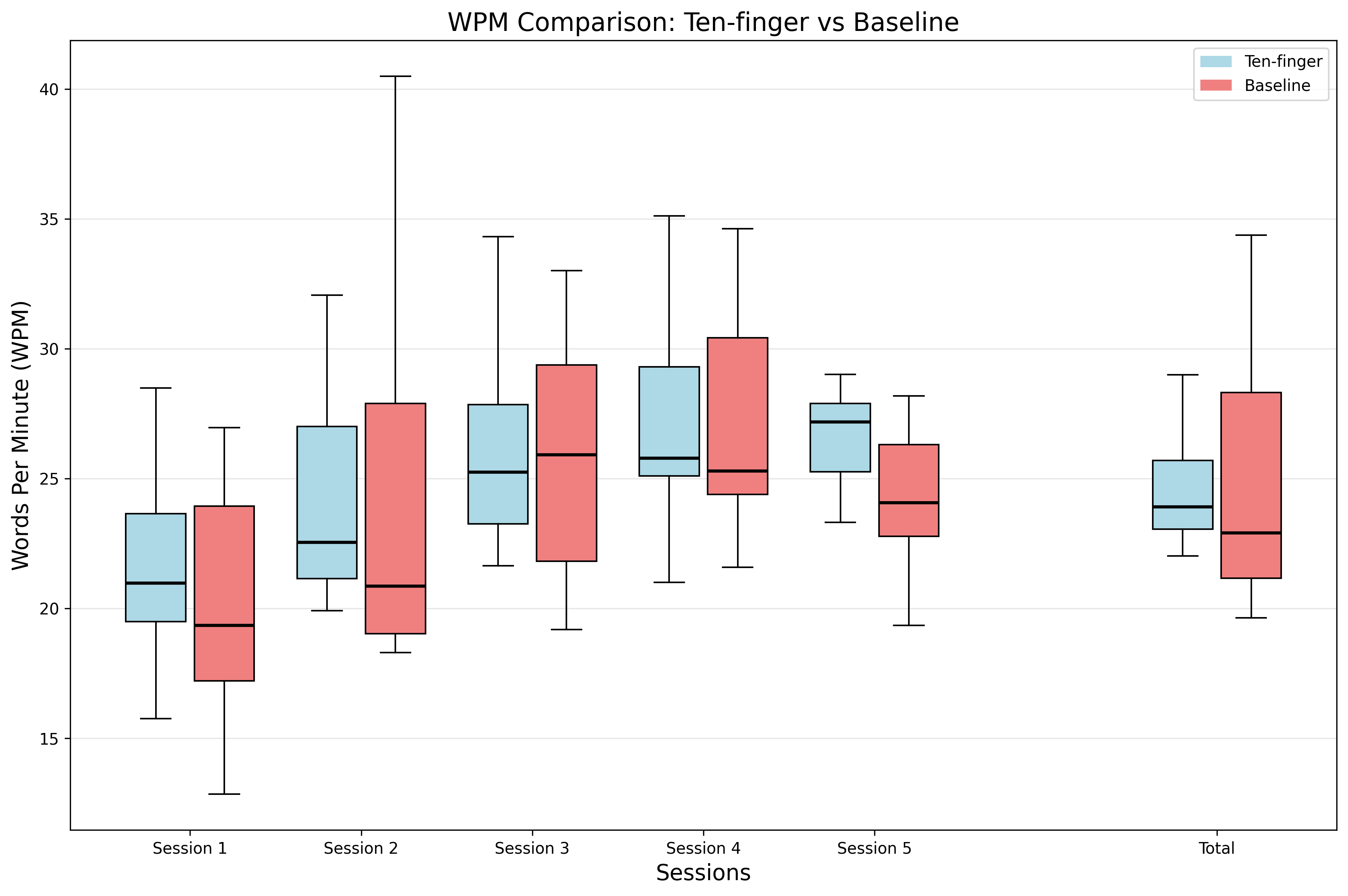}
  \caption{Session-wise WPM for \emph{Ten-finger} (KeySense) and the \emph{Baseline} soft keyboard. Boxes show medians and interquartile ranges; whiskers indicate the data range. \textsf{Total} pools all sessions. Ten-finger speeds increase steadily across sessions, whereas baseline speeds peak mid-study and decline by Session~5.}
  \label{fig:wpm}
  \Description{Session-wise words per minute (WPM) for KeySense vs. baseline. Box-and-whisker plots show KeySense improving across sessions with a pooled median around 25.63 WPM; the baseline is ≈24.73 WPM, has wider dispersion, and declines in the final session.}
\end{figure}

Figure~\ref{fig:wpm} shows session-wise WPM. With Ten-finger, participants exhibited a clear learning effect: a linear mixed-effects model (LMM) on sentence-level WPM estimates a mean gain of 5.12 WPM from Session~1 to Session~5 ($F(1, 223.06)=57.90, p<10^{-12}$). In contrast, baseline speeds peaked around Session~3 and then declined by 1.94 WPM by Session~5 ($F(1, 227)=5.71, p=0.018$), consistent with fatigue on the physically demanding tablet keyboard. By Session~5, Ten-finger was significantly faster than the baseline (28.3 vs.\ 26.2 WPM; $F(1, 118.71)=8.58, p=0.0041$), while overall mean speeds across all sessions remained similar (Ten-finger: $25.63\pm4.23$ WPM; Baseline: $24.73\pm4.67$ WPM) due to initial learning cost. Hands-down ten-finger typing yields a clear speed benefit once users adapt.

\subsubsection{Errors and corrections}

We measure accuracy using Word Error Rate (WER), the word-level minimum string distance normalized by the target length:
\begin{equation}
\mathrm{WER} = \frac{\mathrm{MSD}(S_{transcribed}, S_{presented})}{N},
\end{equation}
where $\mathrm{MSD}$ counts word insertions, deletions, and substitutions, and $N$ is the number of words in the presented sentence. Error rates were low for both conditions: ten-finger averages \(1.20\%\) WER (SD=\(1.75\%\)) and the baseline \(0.85\%\) (SD=\(1.37\%\)); a paired t-test shows the gap is not statistically significant ($t(11)=1.38, p=0.20$).

\tl{To probe correction effort, we use Corrected Error Rate (CER, proportion of words that were initially mistyped but later corrected with \textsf{Backspace}) and correction frequency (corrections per minute). Baseline typing yielded a CER of $3.23\%$ ($\mathrm{SD}=2.47$) and KeySense $4.60\%$ ($\mathrm{SD}=3.06$), again with no significant difference ($t(11)=1.56, p=0.148$). Correction frequency shows a similar pattern (Baseline: $1.77\pm1.19$ per minute; KeySense: $2.03\pm0.95$; $t(11)=0.94$, $p=0.369$). Ten-finger delivers higher speed without imposing a correction burden.}

\subsubsection{Ten-finger behavior and user experience}
\label{subjective}

To understand how people actually type with hands-down contact, we combine two views: behavioral signals from touch logs and subjective ratings from post-study questionnaires.

We first quantify how many touches our pre-decoder treats as intentional. For each word, we compute an intent ratio
\begin{equation}
P = \tfrac{1}{2}\frac{\text{intended threads}}{\text{total threads}}
  + \tfrac{1}{2}\frac{\text{intended events}}{\text{total events}},
\label{eq:intent-ratio}
\end{equation}
which captures both the number of finger threads and the raw touch events interpreted as intentional.

\begin{figure*}[t]
  \centering
  \includegraphics[width=\linewidth]{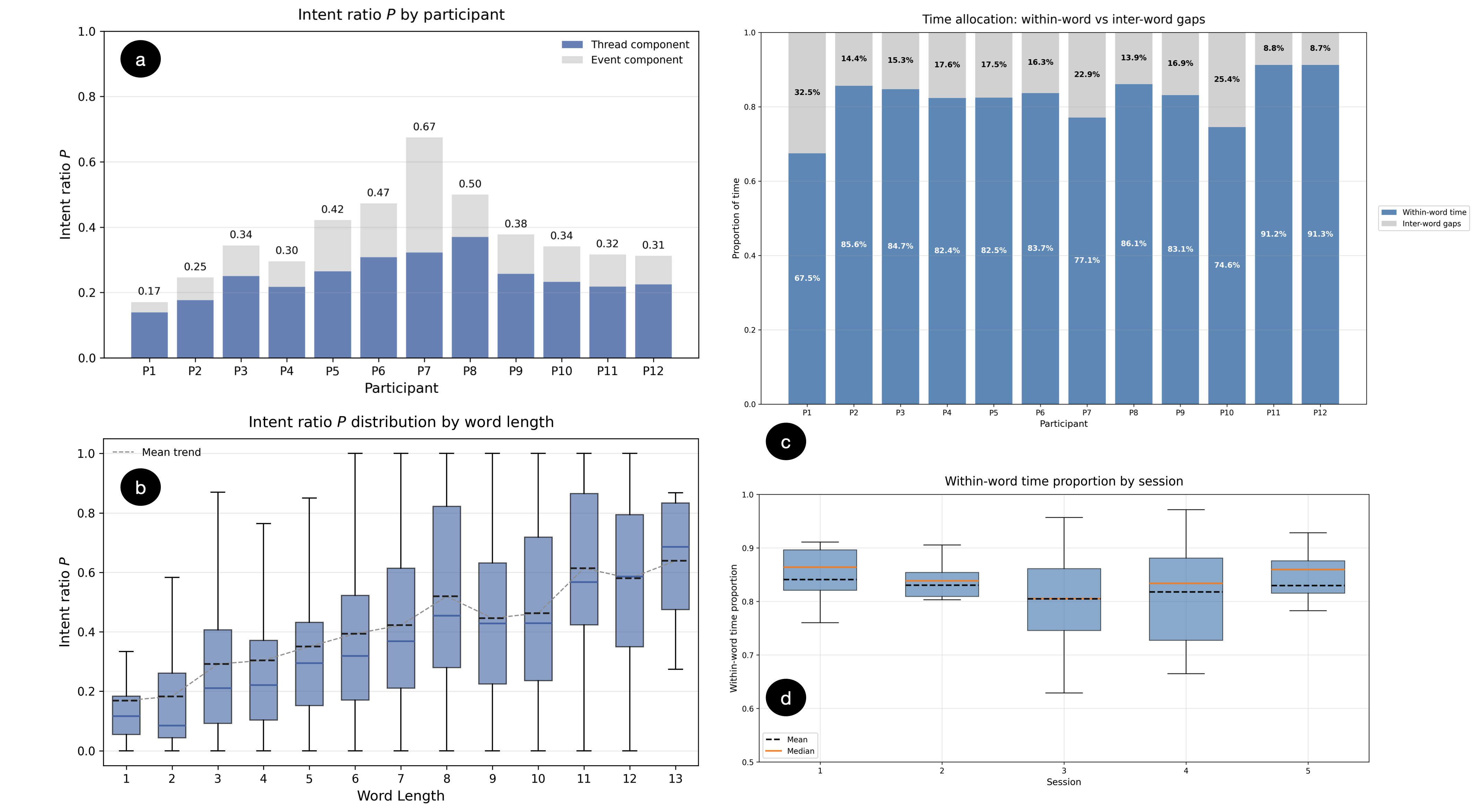}
  \caption{\textbf{Ten-finger typing behavior.}
  \textbf{(a)} Intent ratio $P$ by participant, decomposed into thread and event components; labels show totals.  
  \textbf{(b)} Distribution of $P$ by word length: longer words show higher $P$ as resting fingers stabilize.  
  \textbf{(c)} Time allocation by participant: proportion of time spent within words vs.\ inter-word gaps.  
  \textbf{(d)} Within-word time proportion by session: distributions remain stable.}
  \label{fig:time-intention}
  \Description{Ten-finger typing behavior and time allocation. (a) Intent ratio (P) by participant, shown as stacked bars decomposing P into thread-level and event-level components; numeric labels indicate totals. (b) Distribution of intent ratio P by word length, showing higher P for longer words as resting fingers stabilize and incidental contacts diminish. (c) Per-participant time allocation between within-word typing and inter-word gaps; stacked proportions sum to 1. (d) Distribution of within-word time proportion across sessions, showing that approximately 82.5\% of time is spent within words, with a stable or slightly increasing trend over sessions.}
\end{figure*}

Across all words, the mean \(P\) is \textbf{0.367} (\(\mathrm{SD}=0.296\)), confirming that incidental contacts are frequent but effectively filtered. Panel~\ref{fig:time-intention}a shows large between-user differences, reflecting variations in resting stability and hand posture. \tl{Panel~\ref{fig:time-intention}b shows that \(P\) increases with word length: shorter words often include a brief settling phase with extra contacts, while longer words allow the hand to stabilize as only the reaching finger moves.} Time allocation further reveals a consistent hands-down rhythm (Panels~\ref{fig:time-intention}c--d). Users spent on average \textbf{82.5\%} of time within words (\(\mathrm{SD}=14.4\%\)), with most participants exceeding 80\%. Across sessions, the within-word time proportion is stable or slightly rising and becomes less variable, indicating that users quickly adopt a relaxed rest-and-reach pattern.

\begin{table*}[h]
\centering
\begin{tabular}{|l|l|c|}
\hline
\textbf{Question} & \textbf{Scale 1 -- 5} & \textbf{Median Score} \\
\hline
How do you like it? & very dislike -- very like & 5 \\
I think I would use it frequently. & strongly disagree -- strongly agree & 4.5 \\
It improves conventional typing. & strongly disagree -- strongly agree & 4 \\
Similarity to physical typing & very different -- very similar & 4 \\
Mental demand (Conventional vs. Ten Finger) & very low -- very high & 1 vs. 2.5 \\
Physical demand (Conventional vs. Ten Finger) & very low -- very high & 4 vs. 1.5 \\
\hline
\end{tabular}
\caption{Summary of median subjective feedback scores.}
\label{tab:feedback}
\Description{Summary of subjective measures (medians). Preference = 5; “would use frequently” = 4.5; “improves conventional typing” = 4; “similarity to physical typing” = 4; NASA-TLX mental demand—conventional = 1, ten-finger = 2.5; physical demand—conventional = 4, ten-finger = 1.5.}
\end{table*}

Subjective ratings closely mirror these behavioral patterns. \tl{We adopted a subset of NASA-TLX items \cite{hart2006nasa} using a simplified 5-point version of the original scale (1=very low, 5=very high).} Participants rated the ten-finger method highly (median liking = \textbf{5}; frequent-use intention = \textbf{4.5}). They agreed it improves conventional typing and feels similar to physical typing (both medians = \textbf{4}), suggesting successful transfer of muscle memory. For workload, ten-finger required slightly more mental attention at first (median 2.5 vs.\ 1.0), as participants broke the long-standing habit of avoiding incidental touches, but substantially reduced physical demand (1.5 vs.\ 4.0).

Open-ended comments reinforce this picture. \textbf{P3} notes, ``The 10-fingers virtual keyboard performs much better for long words and feels smoother for short words since I don’t need to press keys down.'' \textbf{P1} writes, ``I’m used to keeping my fingers in the air when typing, so it’s hard to switch to resting on the screen. But if virtual keyboards become common, this could really matter.'' \textbf{P9} summarizes, ``Pros: easy to press a key, no need to push hard. Cons: no physical feedback---need a bit more attention to avoid mistakes.'' Collectively, these results show that users quickly settle into a stable rest-and-reach rhythm with clear ergonomic benefits.

\subsubsection{Decoding latency}
\label{usability}


\tl{To check whether remote LLM decoding could be a hidden bottleneck, we also measured end-to-end latency. \xj{The latency was short and unlikely to affect the input performance.} Decoding requests were sent from the iPad interface to a remote LLM server via an encrypted tunnel (\textsc{Ngrok}) \cite{ngrok2014}. LLM inference takes \textbf{53.70\,ms} (SD=11.94ms). Outbound transmission averaged \textbf{31\,ms} (95th percentile \textbf{40\,ms}), and inbound transmission averaged \textbf{25\,ms} (95th percentile \textbf{29\,ms}). This yields an average round-trip communication delay of roughly \textbf{56\,ms} per decoding request, well below typical thresholds for perceivable lag in visual feedback. The end-to-end \textbf{110\,ms} delay is 55\% lower than the existing cloud-based decoder \cite{ma2025llm}. No participants reported hesitation or delay in suggestion appearance or word confirmation, suggesting that latency did not measurably affect perceived responsiveness.}


\section{Discussion}

\xj{\subsection{Initial barriers and long-term potential} 
By the end of the study, participants achieved an average speed of 28.6 WPM on KeySense, which was similar to other ten-finger typing methods (e.g., 30 WPM on ResType \cite{li2023restype} on the first day). Nonetheless, this performance was still slower than the typing speeds previously reported for the built-in iPad keyboard in prior literature \cite{li20111line}. A primary constraint was participants’ unfamiliarity with the hands-down, ten-finger typing mode. While placing both hands on the tablet surface increased physical comfort, this unfamiliar input style required users to unlearn the habit of hovering their hands above the keyboard and to develop confidence that resting both hands would not trigger false touches. This additional learning demand elevated cognitive load and initially hindered performance, as indicated by subjective ratings and user feedback. Furthermore, only 2 participants in our study identified themselves as typing experts, whereas in the prior work~\cite{li20111line}, all participants were expert typists with an average Qwerty speed of 82.6 WPM.

Despite the initial learning cost, the hands-down ten-finger typing is promising and may offer substantial long-term benefits. Typing speed improved steadily across the five sessions in our study, and the greatest improvement over the hover baseline occurred at the last session. This result also echoes earlier findings that ten-finger typing performance increases rapidly with practice~\cite{li2023restype}.  As it reduces physical demands, alleviates hand fatigue, and more closely resembles the ergonomics of physical-keyboard typing, it has strong potential to become a widely adopted typing mode for tablet-scale touch surfaces.

To fully realize the potential of hands-down ten-finger typing, smoother motor transfer should be achieved. \tl{In our current design, users must frequently shift visual attention between the text area and the keyboard region, because virtual keys lack the haptic cues and home-row bumps of physical keyboards.} Future work could ease this transition by \tl{adding visual anchoring cues that indicate home-row position within the main work area, which reduces gaze switching and makes users focus entirely on the text}. Structured onboarding tutorials and practice curricula could also help users reframe the tablet surface as a “restable” keyboard.}

\subsection{LLM Architecture Affects Decoding Accuracy}
\label{subsec:arch-matters}

Section~\ref{sec:synthetic-decoding} showed that a fine-tuned FLAN--T5-small can substantially outperform carefully engineered Bayesian decoders. A natural question is whether this advantage is specific to our chosen model family or would arise with any sufficiently large LLM. To probe this, we ran a small zero-shot comparison across several open LLM families on easy tasks.

A qualitative pattern emerged: encoder--decoder models from the T5 family behaved markedly better than decoder-only models when asked to emit a single corrected word. In our runs, Qwen-7B \cite{qwen2023qwen}, Gemma-7B \cite{gemma2024}, and Llama-3.1-8B \cite{llama3_2024} (all decoder-only) produced low accuracy, while FLAN--T5 variants ranked higher despite far fewer parameters. This suggests that for noisy-to-clean, single-word normalization, architectural fit matters at least as much as raw scale.
\paragraph*{Why architecture matters.}
\textbf{Cross-attention fits ``map-and-say'' tasks.}
T5-style models first encode the noisy sequence and then let the decoder attend over a fixed representation, closely matching our ``map-and-say'' objective. This separation reduces copy bias and instruction leakage: the prompt and target are distinct streams, and the decoder can focus on reconstructing the word. Decoder-only models instead treat prompt and target as a single left-to-right sequence; when the desired output is short, their strong language priors over the prompt often dominate, yielding safe, prompt-like words instead of strict spellings.

\textbf{Instruction tuning aligns behavior for concise outputs.}
FLAN variants are instruction-tuned and, even without task-specific fine-tuning, follow constraints like ``output only one word'' more reliably than general decoder-only baselines. However, without domain-specific fine-tuning they still lack a robust ``noisy to clean spelling'' mapping, which explains their modest zero-shot performance.

\textbf{Scale helps less than match.}
Increasing parameter count in decoder-only models did not close the gap with T5 on this task: larger variants continued to exhibit prompt-following and over-generation behaviors. By contrast, scaling within the same family (e.g., from FLAN--T5-small to FLAN--T5-XL) improved zero-shot accuracy, suggesting that for spelling-like normalization the family match is more important than raw size across architectures.

These observations reinforce our conclusion: architectural alignment to the problem class matters. A compact fine-tuned FLAN--T5-small can excel while remaining small enough to support real-time use and flexible deployment.

\subsection{Cloud vs.\ On-device Decoder Implementation}

A practical keyboard decoder must keep latency low, protect privacy, and run across heterogeneous devices. \tl{In our setting, the latency constraint is particularly strict: ten-finger typing relies on pre-decoding after each touch and delivers real-time visual feedback, so any additional delay would be directly perceived as lag.} At the same time, we aim to support commodity tablets today and in the longer term, other passive surfaces and form factors.

On-device inference offers stable low latency, full offline privacy, and the opportunity for user-specific adaptation (e.g., private LoRA \cite{hu2022lora} adapters). In our case, a quantized FLAN--T5-small (77M parameters) can plausibly fit on modern tablets and phones (roughly 100\,MB in int8) and matches our single-word inputs well. The drawbacks are energy cost on mobile CPUs/NPUs, tighter memory limits that preclude larger checkpoints, and fragmented hardware/software stacks that complicate maintenance. One plausible deployment profile is therefore a pure on-device mode for privacy-sensitive or offline settings, at the cost of using a smaller model and additional engineering for hardware-specific optimization.

Cloud inference, in contrast, trades network latency for accuracy headroom and operational agility. Hosting the decoder on a server allows us to leverage larger or rapidly updated checkpoints while maintaining uniform behavior across devices. \tl{In our deployment, the additional end-to-end delay introduced by cloud decoding averaged about 56\,ms, well below perceptual thresholds for real-time visual feedback and consistent with prior on-cloud decoders \cite{ma2025llm,zhao2025tap}.} To mitigate privacy and bandwidth concerns, our pipeline keeps all pre-decoding on-device (threading, time clustering and bounded-edit candidate generation) and uploads only a compact package: the noisy letter sequence, not raw touch logs.

Given these trade-offs and our goal of robust accuracy under highly ambiguous inputs, our current implementation deploys the LLM decoder in the cloud while keeping sensing and pre-decoding local. \tl{This split---heavy, privacy-sensitive processing on-device; compact symbolic queries to a neural backend---offers a flexible pattern for LLM-powered text entry: the user interface and touch-processing pipeline remain stable, while the decoder itself can evolve over time or migrate between cloud and on-device execution as application constraints demand.}

\subsection{Limitations and Future Work}

Our study deliberately targets a slice of the design space: word-level correction on a fixed QWERTY layout, in English. This yields clear attributions for our decoder comparison but also omits several factors that matter in everyday typing: multi-word context, out-of-vocabulary content (e.g., names, slang, mixed-language phrases), more extreme or bursty errors, and long-term use in the wild. We highlight four concrete next steps:

\textbf{From word-level to context-aware decoding.}
All decoders in this work operate at the word level and ignore sentence context, even though longer-range semantics could help disambiguate difficult cases and stabilize corrections over time. A natural extension is to lift the LLM decoder to sentence- or phrase-level operation. This would bring hands-down typing closer to continuous text entry, and allow future work to study how context-sensitive correction interacts with user trust, error visibility, and learning.

\textbf{Handling OOV words.}
\xj{Similar to other intelligent decoders, an LLM-based decoder also struggles with out-of-vocabulary (OOV) words, especially when they appear for the first time. One potential solution is to preserve the literal string, the exact characters typed by the user, on the suggestion strip as a viable fallback before applying auto-correction. This allows users to select it in case a false correction is made. Moving forward, once the decoder correctly identifies an OOV word, it should incorporate that word into its dictionary to support smoother and more accurate future input.}


\textbf{Hybrid decoding and longer-term evaluation.}
\tl{Our results show a clear pattern: the statistical decoders perform strongly on very short words, while the LLM dominates on medium and long ones. A promising avenue is to explore hybrid decoders that route inputs based on length or uncertainty: using the statistical decoder for simple short words and the LLM for ambiguous or longer ones. Such hybrids could improve overall accuracy and reduce computation for easy cases. Finally, our five-session user study captures only the early phase of skill acquisition; a longitudinal, in-the-wild deployment would allow us to test these hybrid strategies under realistic conditions and to measure how comfort, speed, and correction patterns evolve over days or weeks.}

\textbf{Cross-user variability and fixed timing.}
\xj{As users' typing behavior varies, personalization could further improve the performance.} For example, our 100\,ms cognitive window was chosen from prior data and works well on average, but individual traces show substantial variation. A static $\tau_c$ therefore limits how well the system can match each user’s rhythm, hinting at further speed and comfort headroom with online personalization of temporal and spatial priors.

Together, these directions move from a controlled prototype toward a context-aware, coverage-robust, and field-tested system that can support high-fluency ten-finger typing on everyday surfaces.
\section{Conclusion}




We set out to enable comfortable, ten-finger typing on commodity, pressure-insensitive touchscreens without adding new hardware. KeySense addresses this challenge by combining a lightweight pre-decoder that interprets hands-down multi-touch input with a task-specialized FLAN--T5-small decoder trained on human-informed synthetic errors, yielding robust word-level corrections from noisy letter sequences. Our component evaluations and user study suggest that this combination is already practically viable: the LLM decoder substantially outperforms statistical baselines and participants quickly attain higher speeds with lower physical demand than conventional hover typing. More broadly, KeySense treats resting-finger contacts as structured signal rather than nuisance, pointing toward ubiquitous, surface-agnostic typing—where any flat surface can serve as a high-performance keyboard, combining ergonomic comfort with competitive speed.

\begin{acks}
We thank the anonymous reviewers for their insightful feedback.
\end{acks}

\bibliographystyle{ACM-Reference-Format}
\bibliography{reference}










\end{document}